\documentclass[11pt, letterpaper]{article}
\usepackage[english]{babel}
\usepackage[utf8]{inputenc}
\usepackage[margin=1in, bottom=1in, top=1in]{geometry}
\pdfpageheight=\paperheight

\usepackage{fancyhdr}
\usepackage{graphicx, enumerate, amsmath, amsfonts, amssymb, amsthm, amsrefs, hyperref, braket}
\usepackage{listings}
\usepackage{caption, wrapfig, subfig}
\usepackage{makecell}
\usepackage{booktabs}
\usepackage{xcolor} 
\title{Temperature nonuniformity Part I}
\date{June 2020}

\graphicspath{{nonuniformity_figures/"}}

\addto\captionsenglish{}
\def\equationautorefname~#1\null{Equation (#1)\null}

\begin{document}
{\Large
\textbf\newline{Resolving nonuniform temperature distributions with single-beam absorption spectroscopy. Part I: Theoretical capabilities and limitations}
}
\newline
\\
Nathan A Malarich\textsuperscript{1,*},
Gregory B Rieker\textsuperscript{1},
\\
\bigskip
1 Mechanical Engineering, University of Colorado at Boulder
\\
\bigskip
* nathan.malarich@colorado.edu

\begin{abstract}
Absorption spectroscopy is traditionally used to determine the average gas temperature and species concentration along the laser line-of-sight by measuring the magnitude of two or more absorption transitions with different temperature dependence. Previous work has shown that the nonlinear temperature dependence of the absorption strength of each transition, set by the lower-state energy, $E^{\prime\prime}$, can be used to infer temperature variations along the laser line-of-sight. In principle, measuring more absorption transitions with broader bandwidth light sources improves the ability to resolve temperature variations.  Here, we introduce a singular value decomposition framework in order to explore the theoretical limits to resolving temperature distributions with single-beam line-of-sight absorption measurements.  We show that in the absence of measurement noise or error, only the first $\sim$14 well-selected absorption features improve the temperature resolution, and a Tikhonov regularization method improves the accuracy of the temperature inversion, particularly for recovery of the maximum gas temperature along the laser beam. We use inversion simulations to demonstrate that one can resolve a selection of temperature distributions along a laser beam line-of-sight to within 3\% for the sample cases analyzed. In part II of this work, we explore the influence of measurement noise and error, and experimentally demonstrate the technique to show that there is benefit to measuring additional absorption transitions under real conditions.
\end{abstract}

\textsuperscript{\textcopyright}  2020. This manuscript version is made available under the CC-BY-NC-ND 4.0 license http://creativecommons.org/licenses/by-nc-nd/4.0/

\section{Introduction}
The performance of many combustion devices is limited by temperature nonuniformity.
In gas turbine engines, for example, so-called ‘hot streaks’ across the combustor exit force operation of engines at sub-optimal operating conditions in order to keep engine components well below material limitations \protect\cites{aero, hotstreak1}.
To improve system efficiencies, several studies have probed the magnitude of this exhaust temperature nonuniformity \protect\cites{hotstreak1, turbine1, turbine2, hotstreak, doll}.
Nonuniformities also impair industrial combustion processes; for instance, temperature nonuniformities introduce unwanted material variability into temperature-dependent flame treatments \protect\cites{flametreatment}.
Real-time control \protect\cites{aero, palaghita} and design of next-generation combustors \protect\cites{hotstreak, doll} would benefit from a technique capable of recovering spatial nonuniformities within these industrial power and manufacturing systems.

\begin{figure}[h]
\centering
\includegraphics[width=3in]{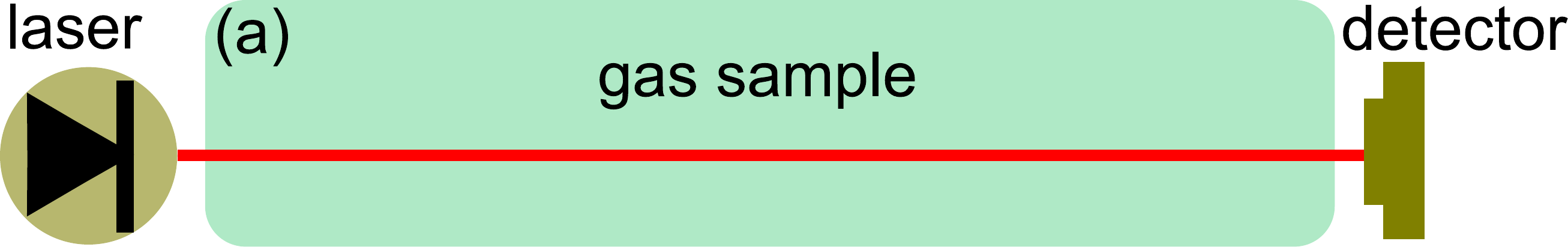}
\vspace{1mm} 

\includegraphics{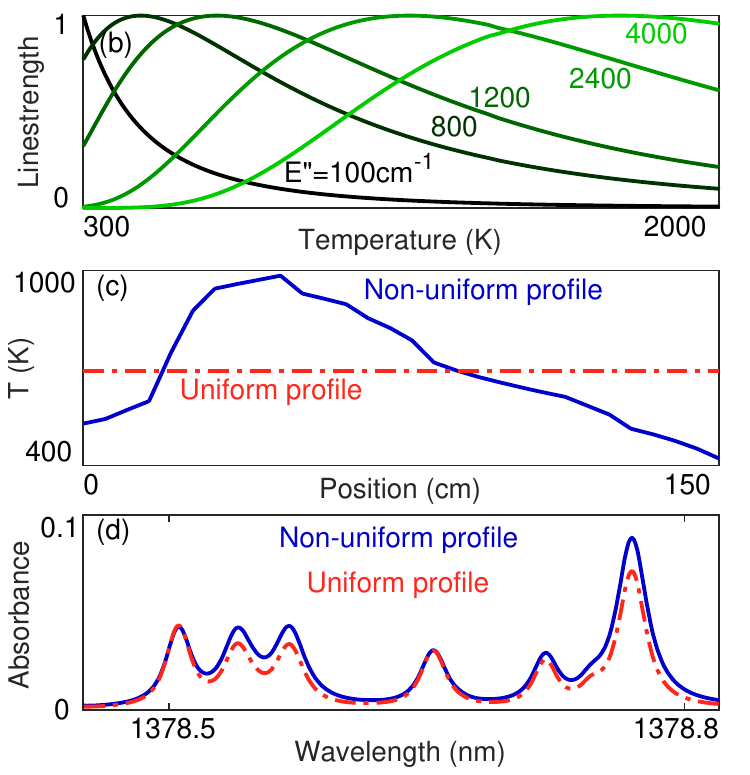}
\captionsetup{width=.5\linewidth}
\caption{a) Line-of-sight absorption spectroscopy schematic. b) Nonlinear temperature-dependence of the linestrength of individual absorption transitions with different lower state energies. c) Uniform and non-uniform line-of-sight gas temperature profiles with same $\overline{T}$. d) Simulated line-of-sight absorption spectra for each of the temperature profiles in (c), assuming uniform $\chi_{H_2O}$ =1\% and P = 635 Torr.}
\label{fig:intro}
\end{figure}

Absorption spectroscopy is a common laser diagnostic technique for providing quantitative thermodynamic data in industrial systems with minimal optical access \protect\cite{goldensteinreview}. In absorption spectroscopy, laser light at a frequency that is resonant with a quantum energy-level transition within an atom or molecule propagates through an unknown environment to a detector (Fig. \ref{fig:intro}a). The amount of laser light that the target species absorbs along the path is directly proportional to the species concentration, and nonlinearly related to the temperature of the system.  The nonlinear response to temperature depends predominantly on the lower state energy ($E^{\prime\prime}$) of the quantum transition, as shown in Fig. \ref{fig:intro}b. A ratio of the absorption on two transitions with different $E^{\prime\prime}$ cancels the linear dependence on concentration and results in a monotonic function of temperature that can be used to measure temperature in a system.  This so-called “two-line” thermometry technique is the most common form of absorption thermometry for gases \protect\cite{2line}. 
      
Two-line absorption thermometry recovers line-of-sight averaged temperature along the laser beam path. A drawback of two-line absorption thermometry is that the mean temperature retrieval is biased by temperature nonuniformities along the laser line-of-sight \protect\cites{goldensteinnonuni, 12}.  This effect stems from the nonlinear temperature dependence of the absorption magnitude of the individual transitions across the spectrum (Fig. \ref{fig:intro}b). Fig. 1c shows a uniform and a non-uniform temperature profile with the same average temperature. Fig. 1d shows the simulated water vapor absorption spectra for these two profiles. Each absorption transition in Fig. \ref{fig:intro}d responds differently to the nonuniformity. Several tactics have been developed to mitigate the bias that this nonlinear temperature sensitivity introduces to the mean temperature retrieval. These tactics include selecting absorption lines with nearly linear temperature dependence in the expected temperature range \protect\cites{goldensteinnonuni, werblinski, blfirst}, extracting temperature from radical chemical species which are only expected to exist in the hottest region of interest \protect\cite{15}, and determining absorption-weighted line-of-sight average temperatures from computational fluid dynamics simulations to make direct experiment/simulation comparisons in nonuniform environments \protect\cite{wms}.  

Perhaps a more tantalizing approach than recovering a non-biased mean temperature is to use the nonlinear response of the various absorption transitions to infer the nonuniform temperature distribution \protect\cite{Sanders}.  We name this approach a temperature inversion because it inverts a matrix relating multiple features of the absorption spectrum to different features of the temperature distribution \protect\cites{Sanders, lbin}. This temperature inversion approach requires an absorption spectrometer that is capable of measuring many transitions in order to resolve more behavior of the temperature distribution. Broad optical bandwidth lasers with the potential to cover 100+ absorption transitions in combustion environments are becoming increasingly ubiquitous, including spectrometers with supercontinuum sources \protect\cites{werblinski, kaminski, vcsel, wagner}, Fabry-Perot and external-cavity tunable diodes \protect\cites{beijing, tomoafrl}, MEMS-VCSEL lasers \protect\cites{Sanders, hayden, vcselafrl, sanders2}, and broadband versions of frequency comb spectrometers \protect\cites{fielddcs, highspeed, gasifier, midirdcs, foltydb, folty_vernier}. It is therefore interesting to consider to what extent current and future broadband laser sources can improve accuracy for nonuniform line-of-sight temperature measurements.

Temperature inversions have been demonstrated with various broadband sources -- some studies measured $\sim$10 absorption transitions \protect\cites{Sanders, lbin, beijing}, while one tomographic study used 100 absorption transitions \protect\cite{tomoafrl}. These studies found that including additional absorption transitions in the analysis improved the accuracy of the recovered temperature distributions. There have been no systematic studies to date on how well single-beam temperature distribution absorption thermometry can resolve the full variety of nonuniform temperature profiles present in combustion.

In this paper, we introduce a new method to formulate the temperature inversion, and use it to define the theoretical capabilities and limitations of absorption spectroscopy to resolve generalized nonuniform temperature distributions. In Section \ref{sec:theory}, we define the inputs and outputs of the inversion as continuous functions, which allows us to leverage truncated singular value decomposition (TSVD) in Section \ref{sec:tsvd}.  TSVD enables estimation of a thermodynamic function from a Fourier series-like approximation of an absorption function. We show with simulation how the TSVD approach can guide how many absorption features are needed for the nonuniform temperature measurement, and why certain temperature profiles are more challenging to recover from absorption spectroscopy. We show why it is overly optimistic to assume that measuring more and more absorption transitions will always improve temperature distribution accuracy.  While the TSVD approach helps define the limitations of the single-beam inversion technique, we show in Section \ref{sec:tikh} that other post-processing algorithms can improve the nonuniform temperature measurement accuracy. Specifically, we demonstrate with simulation how one particular implementation of Tikhonov regularization enables absorption spectroscopy to more accurately recover the more challenging temperature profiles. A companion paper shows the effect of error sources on this nonuniform temperature retrieval and incorporates these algorithms into an experimental case.

\subsection{Defining the inverse problem}
Here, we define the scope of our temperature inversion approach. We restrict to systems with moderate optical depths that follow Beer's Law for absorption of the laser light, which is valid for many measurements in combustion systems. Thus we will omit situations where emission is significant, and the complete radiative transfer equation would be required \protect\cites{exoplanet, ren}. Extension to the complete radiative equation with emission and scattering would complicate but not invalidate our analysis.

Eq. \eqref{eq:beer} shows Beer's Law, written with H\textsubscript{2}O as the absorbing gas species, which is ubiquitous in engineering systems and produces many absorption lines in the near-infrared 6800-7400 cm\textsuperscript{-1}.
\begin{equation}
\label{eq:beer}
\alpha_\nu = \int_0^L P_{H_2O}(x) \sum_{i=1}^{lines} \phi_i(\nu;T,P_j)S_i(T) dx
\end{equation}

$P_{H_2O}$ and $T$ are the spatial profiles of the absorbing gas partial pressure and temperature along the laser beam span of length $L$. The linestrength $S$ and lineshape $\phi$ are both complex nonlinear functions of gas temperature. The subscript $i$ represents an absorption-transition-specific parameter, and $P_j$ is the partial pressure of each gas species.

Temperature dependence of the linestrength $S$ is governed by the lower-state energy $E^{\prime\prime}$ of the rotational-vibrational quantum state of the absorbing molecule in the gas. The nonlinear temperature-dependence in the linestrength is shown in Eq. \eqref{eq:st} and Fig. \ref{fig:intro}b. Here we ignore the stimulated emission term in the linestrength, as this approximation is $>$99\% accurate for near-infrared transitions below 2000 Kelvin.
\begin{equation}
\label{eq:st}
S_i(T; E^{\prime\prime}, S_i(T_0), Q) \approx S_i(T_0) \frac{T_0}{T} \frac{Q(T_0)} {e^{-c_2 E^{\prime\prime}_i/T_0}} \frac{e^{-c_2 E^{\prime\prime}_i/T}} {Q(T)}
\end{equation}

Here $Q$ is the absorbing molecule's partition function, and $c_2$ is a unit-conversion through physical constants $hc / k_B$.
The integrated area of a single absorption line becomes a simpler function of the linestrength $S$ and the partial pressure of the absorbing species $P_{H_2O}$ as shown in Eq. \eqref{eq:path}. The lineshape function is defined to have a frequency-integrated magnitude of unity, so it drops out of the integrated area equation.
\begin{equation}
\label{eq:path}
\int_\nu \alpha_i = \int_0^L P_{H_2O}(x) S_i(T(x))dx
\end{equation}

The integrated area of a water vapor absorption feature is directly proportional to the integrated product of the number of water molecules at each point along the beam and the fraction of those molecules in the lower energy state of the transition at each point. The specific positions of the molecules along the laser beam do not influence the magnitude of the integrated area. Thus Beer’s Law is sensitive only to the temperature \textit{distribution} of absorbing molecules along the laser beam, not the specific temperature \textit{profile} (where \textit{profile} is used to denote the specific arrangement of temperature along the beam path). If however, the gas system has sufficient optical access for multiple intersecting laser beam paths, then tomographic techniques can help recover spatial profiles \protect\cites{tomoafrl, cai_review, grauer}.

The problem of recovering a temperature distribution from an absorption spectrum is an inverse problem because Beer’s Law describes how the unknown temperatures quantitatively determine the known absorption measurement, not vice versa. In principle, any nonlinear temperature-dependence within the Beer’s Law integral (Eq. \eqref{eq:beer}) could be exploited to determine a temperature distribution. This paper considers temperature inversions from the integrated area of the absorption transition because:
\begin{itemize}
  \item The integrated area relates to temperature through the linestrength.
  \item Linestrength is a single parameter that is typically known with lower uncertainty compared to the lineshape parameters \protect\cite{hitran}.
  \item The lineshape depends on the concentration of other species in the absorbing sample, and has a more complex interpretation in a non-uniform environment (e.g. the path-integrated lineshape profile along a non-uniform path departs from the traditional lineshape profiles).  The complexity and uncertainty associated with lineshape models therefore make the lineshape a comparatively poor candidate measurement for inversion problems, where small measurement and model errors can produce large solution errors \protect\cite{invtext1} in the temperature distribution. 
\end{itemize}
This paper explores idealized single-path temperature inversions in order to determine the theoretical limits of the technique. This ideal inversion contains two assumptions:
\begin{itemize}
  \item One has already extracted a set of linestrengths $S$ from the spectrum without error.     
  \item Partial pressure $P_{H_2O}$ is constant but unknown. 
\end{itemize}
In practice, linestrength error and partial pressure nonuniformities introduce additional complexity into the inversion method. These effects degrade the temperature accuracy below the idealized results presented in this paper. The companion paper discusses a new method to extract the linestrengths for non-uniform conditions, and the sources and influence of linestrength error in the context of experimental measurements. The conclusions in this paper about absorption line selection and comparative challenge in retrieving different types of temperature distributions remain valid for real measurements.

\section{Theoretical framework for inversion method}
\label{sec:theory}
In order to optimize the accuracy of the temperature inversion and understand which temperature distribution features (e.g. maximum, minimum, slope, curvature) can be retrieved, we pose three research questions: 
\begin{enumerate}
  \item How many absorption lines are useful to the inversion? 
  \item Which absorption lines are most useful to the inversion? 
  \item How does the shape of the temperature profile affect our ability to resolve the temperature distribution with single-beam thermometry? 
\end{enumerate}
In this section, we cast the inversion problem in terms of continuous functions, beginning with the path-integral (Eq. \eqref{eq:path}). Specifically, Section \ref{sec:pdf} defines the temperature distribution in terms of common probability functions, and Section \ref{sec:shat} defines the absorption transition inputs to the inversion as samples of a continuous spectroscopic function of lower-state energy. By defining this spectroscopic function, we can disentangle the limitations of a particular laser bandwidth from the more fundamental limitations of the temperature nonuniformity inversion technique in general. This function formulation allows us to apply an established matrix inversion method called truncated singular value decomposition (TSVD) to illuminate the three above research questions in Section \ref{sec:tsvd}. We will go on in Section \ref{sec:tikh} to improve on the inversion using Tikhonov regularization, which is ultimately better for solving for temperature distributions (but not for building intuition on the limitations of absorption spectroscopy).

\subsection{Defining the temperature distribution}
\label{sec:pdf}
In this subsection we define the temperature distribution function, which will become the output of the temperature inversion.

As mentioned previously, absorption spectroscopy can determine the temperature distribution, not the temperature profile. 
This distribution appears in a change-of-variables of Beer's Law, where an integrated area is expressed as an integral over gas temperature rather than position along the laser beam.
We show the result of this change-of-variable in Eq. \eqref{eq:inversiondT}, which we derive for general temperature and concentration profiles in Appendix \ref{sec:pdfexact}.

\begin{equation}
\label{eq:inversiondT}
\int_\nu \alpha_i = \overline{P_{H_2O}}L \int_0^\infty TDF(T) S_i(T) dT
\end{equation}
The integrated area $\int \alpha_i$ is measured by absorption spectroscopy, and the linestrength relationship $S_i(T)$ is known from Eq. \eqref{eq:st}.
That leaves as unknowns in Eq. \eqref{eq:inversiondT} the total column concentration of H\textsubscript{2}O molecules, $\overline{P_{H_2O}}L$, and the temperature distribution function, $TDF(T)$. 
Here the line above $\overline{P_{H_2O}}$ indicates the path-averaged quantity.
The temperature distribution (TDF) refers to the fraction of H\textsubscript{2}O molecules from the path of the laser beam that are at temperature $T$.
Note that the TDF bears some similarity to a probability density function (PDF) in random-number statistics.
However, the TDF is based not on random-number probabilities, but instead on the temperature of every H\textsubscript{2}O molecule along one path.

\begin{figure}[h]
 \centering
  \includegraphics{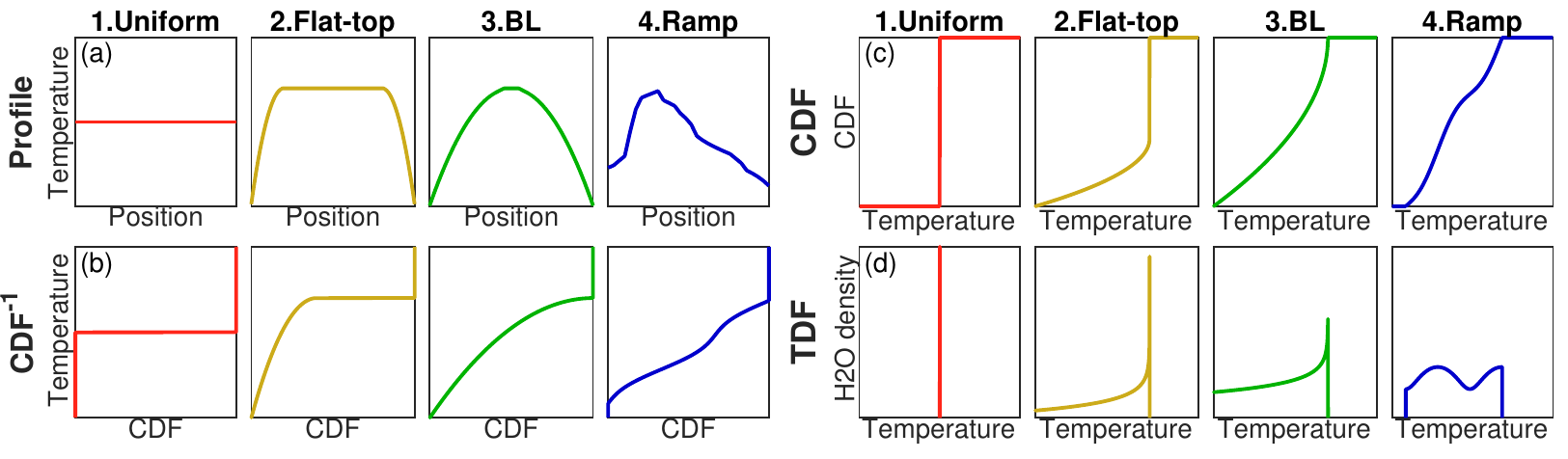}
  \caption{(a) Four diverse temperature profiles used to test the inversion: uniform, developing boundary layer (‘Flat-top’), fully-developed boundary layer ('BL'), and an asymmetric experimental profile measured by thermocouple in 3-zone tube furnace (‘Ramp’). The absorption spectrum is only sensitive to distribution, which can be expressed in three forms: (b) the monotonically-increasing inverse cumulative distribution function (CDF\textsuperscript{-1}) repeated in later figures, (c) the standard CDF, (d) the the temperature distribution function (TDF) describes the fraction of H\textsubscript{2}O molecules from the path of the laser beam that are at temperature $T$. The TDF is on a log scale.}
\label{fig:pdf}
\end{figure}

The temperature distribution function (TDF) can be calculated from any spatial temperature and composition profile.
We show the relationship between a temperature profile and TDF in Fig. \ref{fig:pdf} for the specific case of homogeneous concentration $P_{H_2O}(x)$=$\overline{P_{H_2O}}$.
Four temperature profile test cases are shown in Fig. \ref{fig:pdf}a, representing a variety of different engineering applications. Case 1 shows a uniform hot temperature region spanning the entire laser region. Test Cases 2 and 3 represent a partially or fully-developed boundary layer between, for example, a combustion zone and the colder surroundings. Finally, Test Case 4 demonstrates a more complex asymmetric profile.

We can relate the temperature profiles in Fig. \ref{fig:pdf}a to their corresponding TDFs in Fig. \ref{fig:pdf}d.
First, we sort the temperature profiles in Fig. \ref{fig:pdf}a into monotonic functions in Fig. \ref{fig:pdf}b. 
For the first three symmetric test case profiles, this monotonic function looks like the left half of the profile.
For these homogeneous-concentration profiles, the x-axis of Fig. \ref{fig:pdf}b represents the fraction of H\textsubscript{2}O molecules along the laser path below some temperature $T$.
We label this x-axis "CDF" in reference to the cumulative distribution function $CDF(T)$.
We call this plot the inverse CDF, because it is the same plot as the CDF in Fig. \ref{fig:pdf}c but with flipped x-y axes. We can then relate this CDF to the TDF in Eq. \eqref{eq:pdf}.
In Section 3, we will show the TDF as the output of the matrix inversion, and the inverse CDF as the modified output that looks the most like the temperature profile.
Note that for nonuniform $P_{H_2O}$ concentration profiles, the transformation from profile to inverse CDF becomes more complicated as described in Appendix \ref{sec:pdfexact}.

\begin{equation}
\label{eq:pdf}
TDF(T)= \frac{d(CDF)} {dT}
\end{equation}

All four test-case TDFs show a discontinuity at the maximum temperature, as there must be molecules below but not above $T_{max}$. So $TDF(T)$ is a piecewise continuous function defined over all positive temperatures with discontinuities at $T_{min}$ and $T_{max}$.
We can formalize the definition of $T_{max}$ from the distribution functions in Eq. \eqref{eq:tmax}.

\begin{equation}
\label{eq:tmax}
T_\textrm{max} = \min_{T \in S}\ T, \textrm{ where } S=\left\{T| \int_0^T TDF(x)dx= \int_0^\infty TDF(x)dx \right\}
\end{equation}

Section \ref{sec:tsvd} will show how these discontinuities impair the temperature inversions.

\subsection{Integrated area/Linestrength input to the inversion}
\label{sec:shat}
Just as the temperature distribution solution can be expressed as a continuous function of temperature in Eq. \eqref{eq:inversiondT}, absorption spectroscopy measurements of the integrated area can be normalized into a continuous function of lower-state energy $E^{\prime\prime}$, which we name the normalized linestrength function. Here, we define a function similar to the Boltzmann plot of aerospace and plasma research \protect\cites{boltzplot1, boltzplot2}, and show how temperature nonuniformity changes this function.
In Section \ref{sec:tsvd} we will use a matrix inversion to infer a temperature distribution from this function. In the companion paper, we show how to extract this function from an absorption spectrum. 

\begin{figure}
\centering
\includegraphics{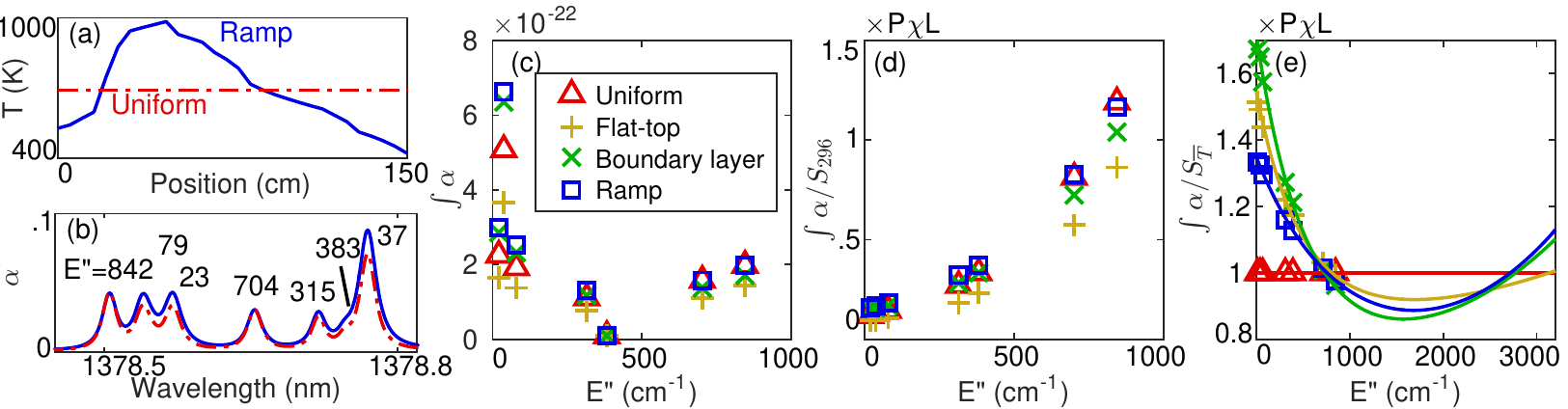}
\caption{Normalized linestrength derivation. (a) Temperature profiles with same mean temperature, from test case 1 and 4, with (b) corresponding absorption spectra. (c)  Seven integrated areas observed in (b) with respect to lower-state energy, extended to all 4 test cases from Fig. \ref{fig:pdf}. (d) Dividing each absorption line area by the transition-specific 296K reference linestrength. (e) Dividing instead by the linestrength at the path-averaged temperature reveals a horizontal line for uniform profiles, and distinct curves for each non-uniform profile. The curves in (e) are not regression fits; they are continuous functions calculated from Eq. \eqref{eq:snorm}.}
\label{fig:snorm}
\end{figure}

Fig. \ref{fig:snorm} shows the derivation of the normalized linestrength function for the temperature profiles in Fig. \ref{fig:pdf}. The simulated absorption measurement shown in Fig. \ref{fig:snorm}b contains seven observable features whose integrated areas change due to the temperature nonuniformity shown in Fig. \ref{fig:snorm}a. In a physical measurement, one would fit the spectra for the integrated areas. For this paper, we calculate the integrated areas from Eq. \eqref{eq:inversiondT} for each of the four test-case temperature distributions, $TDF(T)$, and for each of the seven relevant linestrengths, $S_i(T)$. Fig. \ref{fig:snorm}c presents these integrated area measurements with respect to the lower-state energy, $E^{\prime\prime}$, because the lower-state energy describes the temperature dependence of the linestrength. 

The Fig. \ref{fig:snorm}c scatter plot indicates clear variations in the integrated area due to different temperature profiles, but the complexity in the linestrength $S(T)$ obscures any clear pattern in these variations. This is because the linestrength depends on the absorption-transition-specific lower-state energy $E^{\prime\prime}$ and reference linestrength $S(T_0)$, according to Eq. \eqref{eq:st}.
Thus the transition-specific parameters ($E^{\prime\prime}$ and $S(T_0)$) in conjunction with the measurement-specific TDFs produce the integrated area variations across Fig. 3c. The molecule-specific partition function $Q$ does not cause the linestrength scatter. The lower-state energy and reference linestrength come from a molecular database such as HITRAN \protect\cite{hitran}. HITRAN uses a reference temperature $T_0$ = 296 Kelvin, so the reference linestrength in HITRAN follows the notation $S(T_0=296\textrm{ K}) = S_{296}$.  

After dividing by the HITRAN reference linestrength $S_{296}$ in Fig. 3d, trends in the integrated area for different non-uniform profiles emerge in a visible pattern with respect to lower-state energy, which could be fit with a unique curve for each test case.
Boltzmann plots also divide by the reference linestrength, and then further take the logarithm of this y-axis quantity to determine the mean temperature of the gas \protect\cite{boltzplot1}. In this work we name the y-axis the "normalized linestrength" and do not take the logarithm.

To further clarify the influence of temperature non-uniformity, we change the reference linestrength from the HITRAN reference linestrength to the path-average temperature $S(\overline{T})$, using Eq. \eqref{eq:st}. Then when we divide the integrated areas by this path-averaged reference linestrength, then we get the interesting pattern in Fig.3e. In Fig.3e, the integrated areas for the uniform-path test case collapse to a horizontal line, while the nonuniform test cases follow distinct smooth curves of $E^{\prime\prime}$. This pattern indicates a potential method for distinguishing non-uniform temperature profiles.

We can use any reference linestrength temperature $T_0$ for the matrix inversion input in later sections, so the scatter plots of Fig.3d and Fig.3e are equally useful. This is important, as we do not know $\overline{T}$ \textit{a priori} in a measurement of an unknown environment.  The only significance of choosing $T_0=\overline{T}$ over $T_0=296K$ or any other temperature is that the curve in Fig.3e clearly differentiates between uniform and nonuniform temperature distributions. The normalized linestrengths for the uniform test case (triangle markers) follow a horizontal line in Fig.3e, whereas in Fig.3d they follow the curve $S_i(\overline{T})/S_i(296)$.

We return to the line-of-sight integral Eq. \eqref{eq:inversiondT} to justify the behavior in Fig.3e. By dividing both sides of Eq. \eqref{eq:inversiondT} by a reference linestrength $S(T_0)$, we get two equivalent mathematical formulations of what we call the normalized linestrength, $\hat{S}(E^{\prime\prime})$. The first is derived from a measurement of integrated areas, as Fig. \ref{fig:snorm} illustrates. The second is calculated using the normalized linestrengths from the TDF.
\begin{equation}
\label{eq:snorm}
\hat{S_i}(E^{\prime\prime};T_0) = \underbrace{\frac {\int_\nu \alpha_i} {S_i(T_0)\overline{P_{H_2O}L}}}_\textrm{Derived from $\int\alpha$ measurement}  = \underbrace{\int_0^\infty \frac{S_i(T_j)}{S_i(T_0)} TDF(T_j)dT_j}_\textrm{Calculated from TDF}
\end{equation}
The smooth curves in Fig.3e are not regression curve fits; they are continuous functions calculated from the right-hand side of Eq. \eqref{eq:snorm} for all $E^{\prime\prime}<$3300 cm\textsuperscript{-1}. For the uniform path, $TDF(T_j)$ is nonzero only when the variable of integration $T_j$ equals the path-average temperature, $\overline{T}$. When, as in Fig.3e, we also select the reference linestrength temperature $T_0$ to be the path-average temperature $\overline{T}$, then the $S_i(T_j)/S_i(T_0)$  term becomes equal in numerator and denominator at the only $T_j$ where the TDF is non-zero, so the path-integral collapses to $1$ for all $E^{\prime\prime}$. The nonuniform test cases have components with $TDF(T_j \neq T_0 )>0$, so the non-unity $S_i(T_j)/S_i(T_0)$ term in the integral produces some $E^{\prime\prime}$-dependence in those nonuniform-path normalized linestrength curves. 

We are now prepared to perform matrix inversions of the normalized linestrength function to determine the temperature distribution.  The normalized linestrength curve also has important implications on traditional line-of-sight average temperature measurements using absorption spectroscopy, which are discussed in Appendix \ref{sec:2line}.

\section{TSVD Inversion}
\label{sec:tsvd}
In Eq. \eqref{eq:snorm}, we formulated the nonuniform-temperature inversion as a change of variables from the normalized linestrength curve into the TDF curve. A matrix inversion requires discretization of these input and output curves into vector form – the normalized linestrength measurement according to the specific set of absorption lines and the TDF according to a specified temperature resolution. This discretization changes the normalized linestrength temperature-integral in Eq. \eqref{eq:snorm} into an Einstein-notation matrix expressed in Eq. \eqref{eq:matrix}, with $i$ indicating rows and $j$ indicating the columns and the matrix shown in parentheses.
\begin{equation}
\label{eq:matrix}
\hat{S}(E^{\prime\prime}_i) = \left( \frac{S(E^{\prime\prime}_i,T_j)}{S(E^{\prime\prime}_i,T_0)} \right) TDF(T_j)
\end{equation}
Linear algebra suggests that a laser spanning $m$ different absorption lines with $m$ different $E^{\prime\prime}$ can recover the TDF at up to $m$ different discrete temperatures, provided that each line has unique temperature sensitivity. In this section, we perform a more detailed matrix analysis using singular value decomposition (SVD).  The analysis reveals that the temperature-dependencies of these absorption lines are not fully independent, and therefore adding additional absorption lines provides less information about the TDF than the earlier intuition suggests.  We will also show how the shape of the temperature profile influences our ability to solve for the temperature distribution using single-beam laser absorption spectroscopy.

\subsection{TSVD approach}

Singular value decomposition (SVD) is a generalized form of eigenvalue decomposition for matrix inversions. SVD presents a unique inversion framework in that certain features of the normalized linestrength curve are directly related to corresponding features of the temperature distribution solution. SVD has the advantage of being a fast estimate of the solution from the available measurements. SVD decomposes the matrix in Eq. \eqref{eq:matrix} into two basis vector sets $U_{ik}$ and $V_{jk}$ with corresponding singular values $s_k$ \protect\cite{invtext2}. The shared indices in Eq. \eqref{eq:basis} indicates the measurement basis vector $U_i$ is a function of lower-state energy, and the solution basis vector $V_j$ is a function of temperature. Within this SVD framework, the inversion solution (TDF) becomes a weighted sum of basis vectors $V_k$ whose weights in turn are determined by a scaled-basis vector approximation of the measurement vector $\hat{S}(E^{\prime\prime})$ (Eq. \eqref{eq:basis}). Eq. \eqref{eq:basis} divides the amount of basis vector $U_m$ in the normalized linestrength curve by some scalar weight $\delta_{mk}/s_m$, where $\delta_{mk}$ is the Kronecker delta, to determine the amount of the basis vector $V_k$ in the TDF curve. Thus the shape of these SVD basis vectors (rather than the specific temperature bin selections) fundamentally governs the matrix inversion.

\begin{equation}
\label{eq:basis}
\frac {U_{im}\hat{S}(E^{\prime\prime}_k)\delta_{mk}}{s_m} = V_{jk} TDF(T_j)
\end{equation}

\begin{figure}
 \centering
\includegraphics{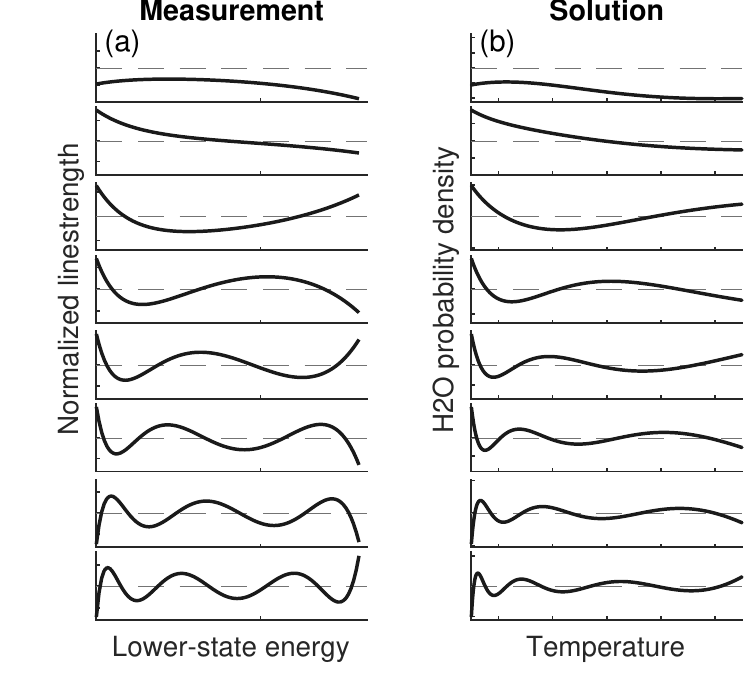}
\captionsetup{width=.5\linewidth}
\caption{First 8 basis vectors from singular value decomposition for (a) normalized linestrength and (b) TDF. Dashed line is at zero.}
\label{fig:basis}
\end{figure}

With adequate lower-state energy and temperature resolution of the matrix, these SVD basis vectors $U_k$ and $V_k$ appear as oscillating functions, similar to the sine and cosine functions of a Fourier series. Shown in Fig. \ref{fig:basis}a, the measurement basis vectors $U_k$ of normalized linestrength are oscillating functions of $E^{\prime\prime}$. Similarly, the solution basis vectors $V_k$ shown in Fig. \ref{fig:basis}b are oscillating functions of temperature. Fig. \ref{fig:basis}a has the same axes as Fig. \ref{fig:snorm}e, and Fig. \ref{fig:basis}b has the same axes as Fig. \ref{fig:pdf}d. The higher-order basis vectors contain more oscillations, in the same manner as higher-order Fourier series terms. The specific shape of these SVD basis vectors is not precisely a Fourier series due to the asymmetric nature of the Boltzmann distribution within Beer's Law, but the oscillation intuition remains the same. As the laser measures more transitions from more $E^{\prime\prime}$ values, the matrix inversion can incorporate more basis vectors to produce a more accurate temperature distribution, subject to measurement precision.

 The inversion method used in this section, truncated singular value decomposition (TSVD), fits the first $p$ singular vectors to the measurement to produce the series solution in Eq. \eqref{eq:tsvd}. The number of usable basis vectors in the inversion stems from the singular values $s_k$ in the matrix decomposition. In Eq. \eqref{eq:tsvd}, the singular values $s_k$ scale the basis vector components of the linestrength measurement to the TDF components.
\begin{equation}
\label{eq:tsvd}
TDF_{TSVD} = \frac {U_{i,k=1:p}\hat{S}(E^{\prime\prime}_i)} {s_k} V_{jk}
\end{equation}

The temperature solution coefficient for Eq. \eqref{eq:tsvd} $V_k$ becomes unresolvable if the uncertainty in the linestrength measurement $\hat{S}(E^{\prime\prime})$ is greater than the singular value $s_k$. The singular values decay to zero for higher-order basis vector terms, so the higher-order terms require a higher measurement accuracy to resolve in the inversion. The matrix rank $p$ in Eq. \eqref{eq:tsvd} describes the number of resolvable basis vectors $V_k$ \protect\cite{invtext2}. For this paper, the uncertainty is governed by the double-floating-point machine error 1e-15, while the lower-state energy selection dictates the decay rate of $s_k$.

\subsection{TSVD results and discussion}
\label{sec:tsvd_results}

\begin{figure}[h]
\includegraphics{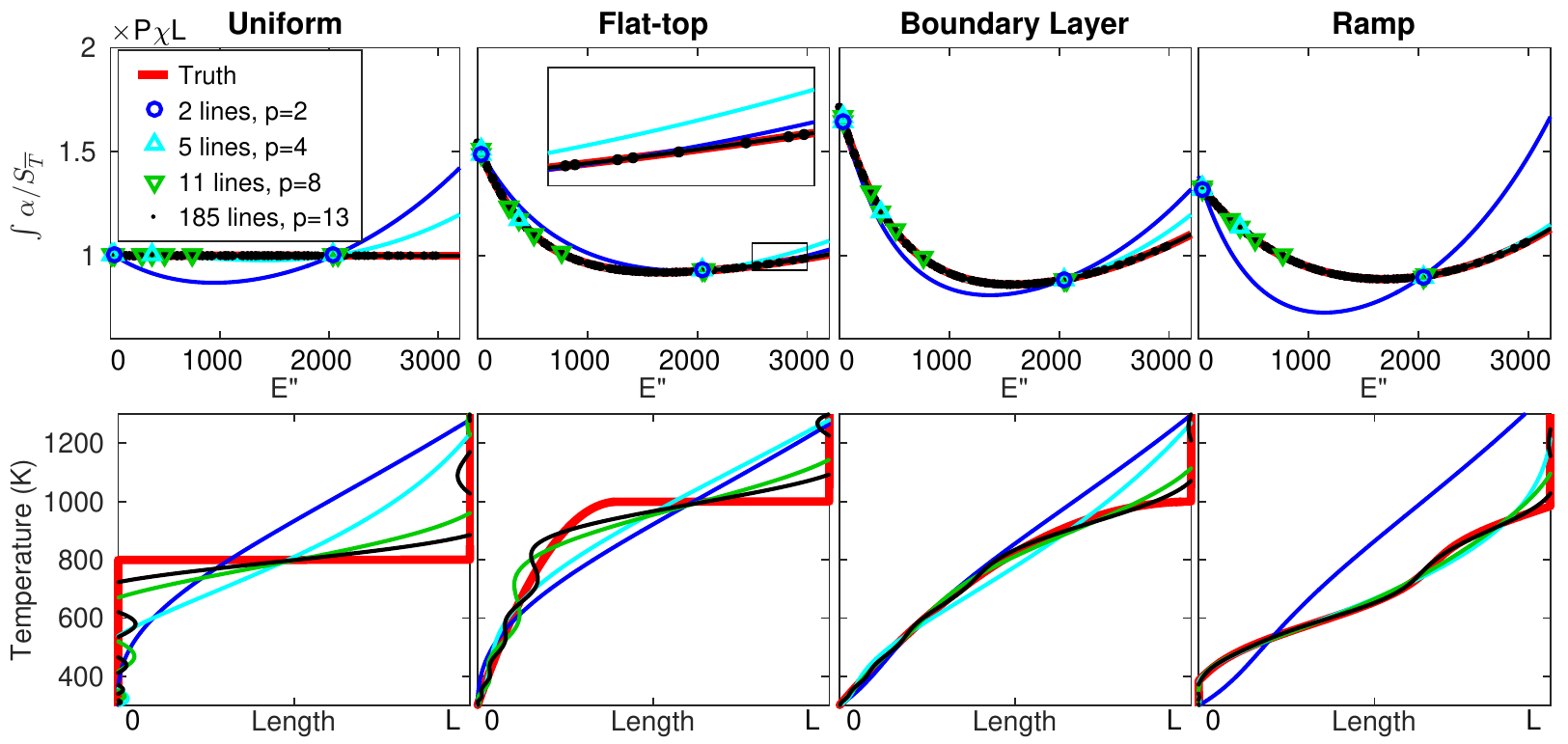}
\caption{Truncated singular value decomposition (TSVD) solutions for four test-case temperature distributions. The markers show where the truth case is sampled for synthetic measurements incorporating various numbers of lines, and the normalized linestrength curves are calculated using Eq. \eqref{eq:snorm} from the TSVD fits. The legend indicates how many $p$ singular values are permitted for each selection of lines (top). Temperature distribution curves (bottom) approach the truth case as $p$ increases.}
\label{fig:tsvd}
\end{figure}

Fig. \ref{fig:tsvd} shows the truncated singular value decomposition (TSVD) inversion method applied to the nonuniform-temperature problem for the four test cases of Fig. \ref{fig:pdf}. The bottom row shows the inverse cumulative distribution functions (CDF\textsuperscript{-1}) of temperature, and the top row shows the corresponding normalized linestrength curves.  The “truth” cases are taken directly from Figs. \ref{fig:pdf}b and \ref{fig:snorm}e, respectively. The markers in Fig. \ref{fig:tsvd}(top) indicate the individual absorption features used for TSVD. They correspond to the lower-state energies of observable water vapor ro-vibrational transitions with $S >$ 1e-23 cm/molecule in an expanding bandwidth region around 6916 cm\textsuperscript{-1}. More lines correspond to a larger optical bandwidth. Because we are ignoring noise for this synthetic measurement, these markers have the same normalized linestrengths as the corresponding points on the truth curve. The “two- line” selection corresponds to the strongest transitions at 6916 cm\textsuperscript{-1}, which would likely be the lines chosen for traditional two-line thermometry measurements. Additional lines are chosen by expanding the optical bandwidth and adding the observable transitions in the new bandwidth. 

The set of lower-state energy transitions from each bandwidth forms a matrix on which we perform SVD to determine the singular vectors and rank $p$. We use these terms in Eq. \eqref{eq:tsvd} to determine the TDF solution, then use Eq. \eqref{eq:pdf} to convert from the TDF to the CDF plotted on the x-axes of Fig. \ref{fig:tsvd}(bottom). In Fig. \ref{fig:tsvd}(top) we show the corresponding full normalized linestrength curves for the TSVD solutions, calculated from the right-hand side of Eq. \eqref{eq:snorm} for all $0<E^{\prime\prime}<$3300 cm\textsuperscript{-1}. The TSVD solutions intersect the truth normalized linestrength curves at each measured $E^{\prime\prime}$, by fitting $p$ singular vectors to the measured linestrengths. To ensure the matrix rank only depends upon the $E^{\prime\prime}$ selection, we use 1 Kelvin resolution over the fit temperature range $T \in [300,1300]$ K.

\subsubsection{How many absorption lines can provide additional accuracy to the inversion and which absorption lines are most useful to the inversion?}

We are now in a position to answer our first two research questions about how many lines, and which lines, are most important to the inversion.  We will see that the answers to these two questions are inextricably linked.

The two-line solutions deviate from the true temperature CDF\textsuperscript{-1} and normalized linestrength curves. However we do expect two-line thermometry to reliably recover the true nonuniform curves. The flat-top normalized linestrength estimation happens to lie near the true curve, while other solutions like the ramp deviate far from the curve for two-line. The first two singular vectors $U_k$ happened to correspond closely with the flat-top distribution. The large deviations in some normalized linestrength curves indicate that those temperature inversions could be improved by adding laser bandwidth and measuring more features of different $E^{\prime\prime}$.

For the five-line measurement in light blue, the normalized linestrength and temperature CDF\textsuperscript{-1} accuracies improve. This result is consistent with previous temperature inversion studies \protect\cites{Sanders, lbin}. Interestingly, although there are 5 lines, the SVD solution can only fit $p = 4$ singular vectors. It turns out two of the lines are a doublet with the same lower-state energy, so the matrix rank corresponds to the number of distinct lower-state energies. The 11-line solution has three repeated $E^{\prime\prime}$, so it has rank 8. The accuracy of the 11-line TSVD solution relies disproportionately on the one line with $E^{\prime\prime} >$1000 cm\textsuperscript{-1}, as otherwise 70\% of the normalized linestrength curve is extrapolated from the data points on the left side of the plot.  \textit{Therefore, the most useful absorption features to a temperature inversion are strong, empirically validated, and distributed over a broad swath of lower-state energy, because these features provide precise, well-spaced points across the normalized linestrength curve.}
  
While initially more absorption lines with different $E^{\prime\prime}$ provide more temperature information, eventually the matrix rank asymptotes. In fact, the 300-1300 Kelvin temperature inversion range in Fig. \ref{fig:tsvd} asymptotes at $p=14$ regardless of how many lower-state energies are measured between $E^{\prime\prime}=0-3300$cm\textsuperscript{-1}.
Using a linear-spaced selection of $E^{\prime\prime}$, we get $p = 14$ using 14 lines and using 3300 lines.
A comparison of the 11-line and 185-line inversions in Fig. \ref{fig:tsvd} illustrates the importance of achieving precise linestrength measurements from the laser absorption spectrum. The difference between the 11-line and 185-line linestrength curve fit is unobservable from the pixel resolution of Fig. \ref{fig:tsvd}(top), but the extra five singular vector terms from the additional high-precision measurements in the 185-line inversion do produce appreciable improvement in the temperature distributions in Fig. \ref{fig:tsvd}(bottom). In turn, the remaining difference between the 185-line temperature distribution and the true temperature distribution in Fig. \ref{fig:tsvd} produces a sub-machine-precision change across the normalized linestrength curve. \textit{Therefore, if one can measure 14 absorption line areas of broadly-selected $E^{\prime\prime}$ to double floating-point precision, no additional temperature information will be gained by measuring additional lines. Our companion paper shows that in real-world measurements with uncertainty, additional laser bandwidth improves the temperature distribution accuracy by reducing the effective uncertainty of the shape of the normalized linestrength curve.} 

\begin{figure}[h]
\centering
\includegraphics{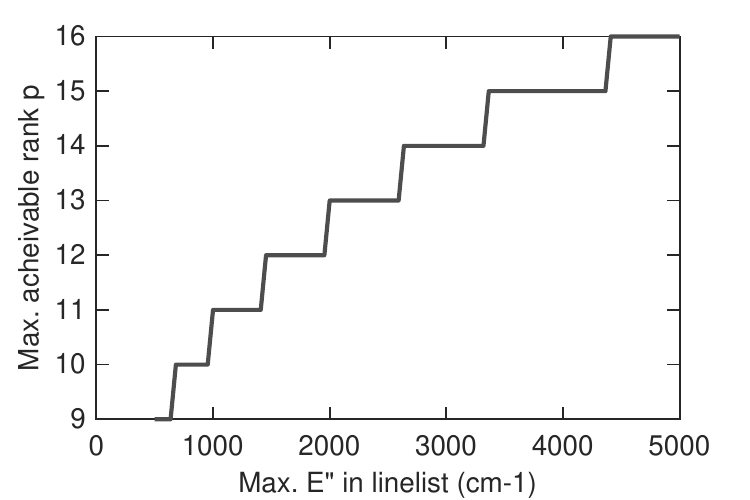}
\captionsetup{width=.5\linewidth}
\caption{Matrix rank from 100 lines selected in different ranges of $E^{\prime\prime}$.}
\label{fig:rank}
\end{figure}

The asymptotic matrix rank depends on the breadth of measured lower-state energies. The matrix rank 14 occurs for $E^{\prime\prime}$ measured between 0-3000 cm\textsuperscript{-1}. The matrix rank remains the same whether there are 185 lines, 3000 lines, or just 14 linearly-spaced lines across this $E^{\prime\prime}$ range. However, the total range of $E^{\prime\prime}$ that is covered by the measurement does matter.  Fig. \ref{fig:rank} shows how the maximum achievable matrix rank increases as the measured range of $E^{\prime\prime}$ increases. The matrix rank increases because the best-fit normalized linestrength curve is defined over all positive lower-state energies. A larger measured range of $E^{\prime\prime}$ constrains more of the normalized linestrength curve for the inversion, requiring less extrapolation. This maximum $E^{\prime\prime}$ is limited by the absorption spectrum selection and noise floor of the laser measurement. Hotter lines with higher $E^{\prime\prime}$ tend to be weaker, because the Boltzmann distribution dilutes over more energy levels at higher temperatures. The highest $E^{\prime\prime}$ feature of the 185-line measurement was located over 100 cm\textsuperscript{-1} away from the location of the 2-line measurement (6916 cm\textsuperscript{-1}). In order to increase the bandwidth of a broadband measurement far enough to get $E^{\prime\prime}_{max} = 3000$, we measured many more absorption features than we would need for a noiseless measurement. A multiplexed laser system would not measure those extra 170 lines. However, if each absorption feature provides an uncertain normalized linestrength, it helps to have the  redundancy of a broad optical bandwidth measurement to better constrain the best-fit normalized linestrength curve.

\subsubsection{How does the shape of the temperature profile affect our ability to resolve the temperature distribution with single-beam thermometry?}
The temperature distribution errors in Fig. \ref{fig:tsvd} are all larger than the diminishingly small normalized linestrength errors. Even the asymptotic $p=14$ solution contains visible temperature distribution errors. In particular, all four test cases systematically overestimate the maximum temperature in the distribution.

We find that $T_{max}$ is difficult to infer from absorption spectroscopy.  This difficulty can be explained using Fourier analysis intuition. Recall from Fig. \ref{fig:basis} that absorption spectroscopy informs the temperature distribution by determining oscillatory basis functions to the TDF. Just as Fourier series require many terms to fit curves around discontinuities, the same is true of the basis functions in this technique. There is a discontinuity in the TDF at $T_{max}$, because $T_{max}$ (1000 K) lies within the temperature inversion search range (300-1300 K), and the TDF is zero above $T_{max}$. The discontinuity in the TDF at $T_{max}$ cannot be resolved with the $p<15$  Fourier-like set of basis functions which are available from a linestrength measurement of any selection of $E^{\prime\prime}$. Accordingly, the uniform and flat-top profiles in Fig. \ref{fig:tsvd}a-b, which have the sharpest discontinuities near $T_{max}$, exhibit the largest inversion errors.
In fact, the uniform and flat-top solutions exhibit visible oscillations above and below $T_{max}$ indicating negative densities of H\textsubscript{2}O ($TDF(T) <$ 0) at some temperatures. 
This negative H\textsubscript{2}O density is an artifact of TSVD, which, in trying to fit a small number of basis vectors to a discontinuity in the distribution, oscillates around the true distribution.
The Ramp test case, whose TDF looks the most like a cosine, converges the fastest of the four test cases, although it also suffers inaccuracies at its $T_{max}$ discontinuity. The consequence of the oscillatory basis vectors is that a feature which requires many basis vectors also requires precise knowledge of the normalized linestrength curve. Many of those basis vectors that are important for the maximum temperature estimation have negligible effect on the normalized linestrength curve. For our test cases, a 1 Kelvin change in the maximum temperature has less than 0.01\% of the influence on the normalized linestrength curve as a 1 Kelvin change in the mean temperature.

We could improve our analysis with more information about the maximum temperature.  For example, if one was certain that the maximum temperature was no higher than 1000 Kelvin (or, say, the adiabatic flame temperature at the operating equivalence ratio), one could set the matrix column of highest temperature near the true $T_{max}$ and recover a more accurate distribution. However, $T_{max}$ is generally the key unknown quantity after $\overline{T}$ \protect\cites{hotstreak1, turbine1}, and a matrix inversion which doesn’t search at temperatures above $T_{max}$ is not actually solving for $T_{max}$. This conundrum led us to pursue more advanced inversion approaches that would explore a broader temperature search range to improve the inversion result in unknown environments. The next section describes this advanced inversion approach.    

\section{Inversion with Tikhonov Regularization}
\label{sec:tikh}
The TSVD inversion method reveals the trends and limitations of a single-beam temperature distribution measurement, but has limitations as a practical approach. Some constraint in the inversion algorithm, in addition to the spectroscopy measurement, is required in order to produce an accurate estimate of the maximum temperature. The general practice of adding constraints to the inversion is known as regularization.  

In this section, we describe a Tikhonov regularization algorithm, which we believe best reconciles the linestrength information about the temperature distribution with our other physical expectations for the system under measurement. This algorithm is substantially the same as used in multi-beam absorption tomography \protect\cite{caima}, which we will motivate with context of the TSVD results in Section \ref{sec:tsvd}, and demonstrate on our four single-beam test-case profiles. 

\subsection{Two inversion constraints}

We first introduce two simple inversion constraints that we will incorporate in Section \ref{sec:limits} into the more complex Tikhonov algorithm: uniform-temperature and non-negative TDF. These two constraints each impose an assumption beyond the spectroscopy information about the physics behind the temperature distribution, but are not by themselves adequate for broadband absorption measurements of nonuniform temperature distributions.

The first constraint, uniform-temperature, has a subtle influence on this problem. It is at first surprising that the uniform temperature distribution has the highest TSVD error of the four test cases in Fig. \ref{fig:tsvd}, whereas traditional absorption thermometry perfectly reproduces only this uniform distribution. In fact, traditional absorption thermometry does not confirm from the spectroscopy whether the profile is uniform, but merely imposes a uniform-profile constraint.  For the two-line solutions in Fig. \ref{fig:tsvd}, this is a reasonable constraint. However, the broadband absorption measurements of additional absorption lines in Fig. \ref{fig:tsvd} provide compelling evidence for nonuniformity in the three nonuniform test cases, and can afford a less restrictive constraint than uniform-temperature.

\begin{figure}[h]
\centering
\includegraphics{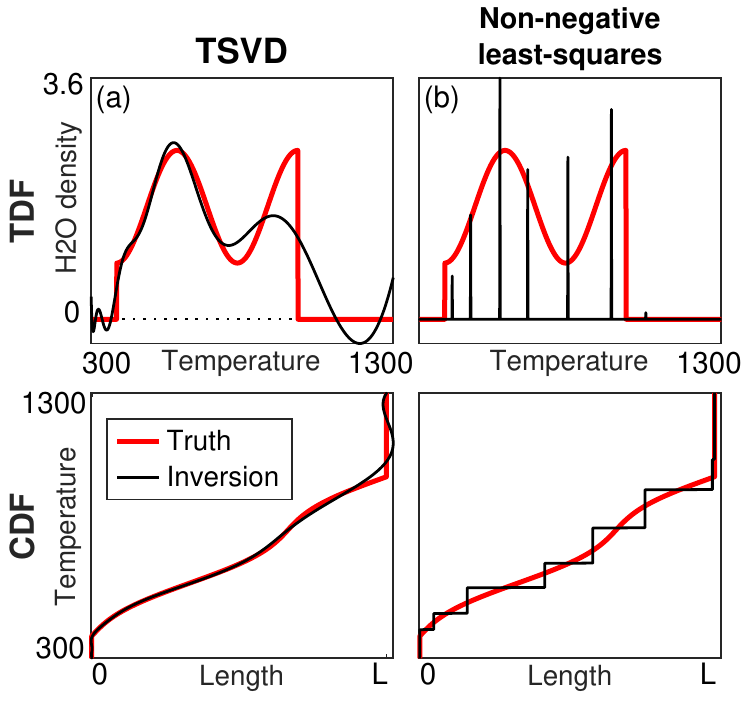}
\captionsetup{width=.5\linewidth}
\caption{Two inversion algorithms each produce nonphysical solutions to Test Case 4. Left: TSVD produces negative H\textsubscript{2}O regions. Right: Nonnegative least-squares produces sharp temperature gradients.}
\label{fig:lsq}
\end{figure}

One less restrictive constraint is the requirement that the TDF be non-negative. This constraint applies the physical law that there cannot be fewer than zero molecules at any temperature. Because a least-squares solution via SVD will fit oscillating functions to the $TDF(T)$ curve, a solution without the non-negative constraint will contain temperatures with negative H\textsubscript{2}O density as in Fig. \ref{fig:lsq}a, particularly around the discontinuity at $T_{max}$. By applying the non-negative constraint, an algorithm must find the most similar solution which cuts out the negative oscillations of the TSVD. Algorithms which use the non-negative constraint, like MATLAB’s $lsqnonneg$ function \protect\cite{lsq}, tend to produce a discontinuous TDF as in Fig. \ref{fig:lsq}b which is no more realistic than the TSVD solution. 

\subsection{Tikhonov Regularization approach}
\label{sec:limits}

Tikhonov regularization, as we define it for this temperature distribution problem, allows a middle ground between the restrictiveness of the uniform-temperature constraint and the nonphysical piecewise solutions of the non-negative constraint. Its use for this temperature inversion problem is not new \protect\cites{caima, beijing, tomoafrl}. Here we describe why this regularization algorithm consistently improves the accuracy of $T_{max}$ for our test case distributions (Fig. \ref{fig:pdf}) as compared with the simpler methods in Fig. \ref{fig:lsq}.  

At its core, Tikhonov regularization adds a second term to the cost function that is minimized by the inversion algorithm. A typical inversion finds the temperature distribution that minimizes the sum-squared error between the measured absorption linestrengths and the $\hat{S}(E^{\prime\prime})$ model. A Tikhonov inversion minimizes this linestrength residual as well as a regularization term which describes the sum-squared gradient in the solution \protect\cite{invtext3}, expressed generically in Eq. \eqref{eq:tikh}. The regularization parameter $\gamma$ determines the relative priorities of minimizing the normalized linestrength residual or minimizing the temperature gradient in the inversion.

\begin{equation}
\label{eq:tikh}
min\left\{ \left\lVert \mathrm{linestrength \ residual} \right\rVert + \gamma \lVert \mathrm{temperature \ gradient} \rVert \right\}
\end{equation}

We want to use the gradient penalty in Eq. \eqref{eq:tikh} to recover the test case temperature distributions in Fig. \ref{fig:pdf}b better than we could using TSVD (Fig. \ref{fig:lsq}a). It turns out the best way to recover these Fig. \ref{fig:pdf} distributions is to change the formulation of the solution vector. In particular, we want regularization to produce a better estimate of $T_{max}$. The traditional formulation of the matrix inversion solves for the TDF in Fig. \ref{fig:pdf}d, where $T_{max}$ is the sharpest, highest-gradient feature on the TDF. In contrast, $T_{max}$ is often located at the smoothest part of the temperature profile and CDF\textsuperscript{-1} (Fig. \ref{fig:pdf}a-b). So the gradient penalty should produce better $T_{max}$ estimates when applied to a temperature profile or CDF\textsuperscript{-1} than a TDF.
This practice of solving for the temperature at several discrete lengths, rather than solving for the TDF directly, is called length-binning \protect\cite{lbin}.
Another advantage of length-binning is that the solution, a discrete set of temperatures present in the gas, necessarily has a non-negative TDF with unity area (see Appendix B, Eq. \eqref{eq:tikhpdf}).

Tikhonov regularization reproduces the test case CDFs the best when penalizing a $2^{nd}$-order temperature gradient on a temperature profile. We want to penalize the temperature distribution gradients most severely at $T_{max}$, because those were the most consistent errors in the TSVD solution (Fig. \ref{fig:tsvd}). TSVD will also produce errors around the discontinuity at $T_{min}$, but because some of our test cases have cold boundary layers, the CDF\textsuperscript{-1} temperature distribution is steeper around $T_{min}$ than $T_{max}$, so the steep TSVD solution is more physically plausible at $T_{min}$.
Our algorithm minimizes the finite-difference gradient $d^2T / dx^2$ to a temperature profile (Eq. \eqref{eq:lbin}), where $T_{max}$ lies towards the center of the profile and $T_{min}$ stays at the edges.
That way, when Tikhonov regularization penalizes temperature gradients on all but the two boundary length bins, the algorithm produces better estimations of $T_{max}$ without artificially increasing $T_{min}$.

\begin{equation}
\label{eq:lbin}
\lVert \mathrm{temperature \ gradient} \rVert^2 \ = \sum_{i=2}^{N_{lengths}-1} \left( T(x_i) - \frac{1}{2} \left(T(x_{i-1}) + T(x_{i+1})  \right) \right)^2
\end{equation}
Eq. \eqref{eq:lbin} is the 2\textsuperscript{nd}-order Tikhonov regularization constraint, first formulated for 3D temperature tomography in \cite{caima} and first written this way for a single-beam problem in \cite{beijing}.

The regularization parameter $\gamma$ in the length-binning solution acts as a lever between the uniform and non-negative least-squares solutions. In the limit $\gamma \to 0$, the temperature gradient term no longer influences Eq. \eqref{eq:tikh}, so the inversion simplifies to a non-negative least-squares solution with implausibly large gradients, as in Fig. \ref{fig:lsq}b. This solution places too much emphasis on the underdetermined linestrength measurement, and inadequate emphasis on our intuition of a realistic temperature profile. From Eq. \eqref{eq:tikh}, in the opposite limit  $\gamma \to \infty$, the temperature gradient term must vanish to achieve a minimum-residual solution. This opposite limit forces the uniform-temperature profile, which is the best single-temperature fit to the linestrength measurement. This uniform-temperature solution places too much emphasis on a minimal gradient restriction, when the linestrength measurements may clearly indicate temperature variations across the laser path. Intermediate values of the regularization parameter produce more accurate solutions in some balance of these limiting cases.

\begin{figure}[h]
\centering
\includegraphics{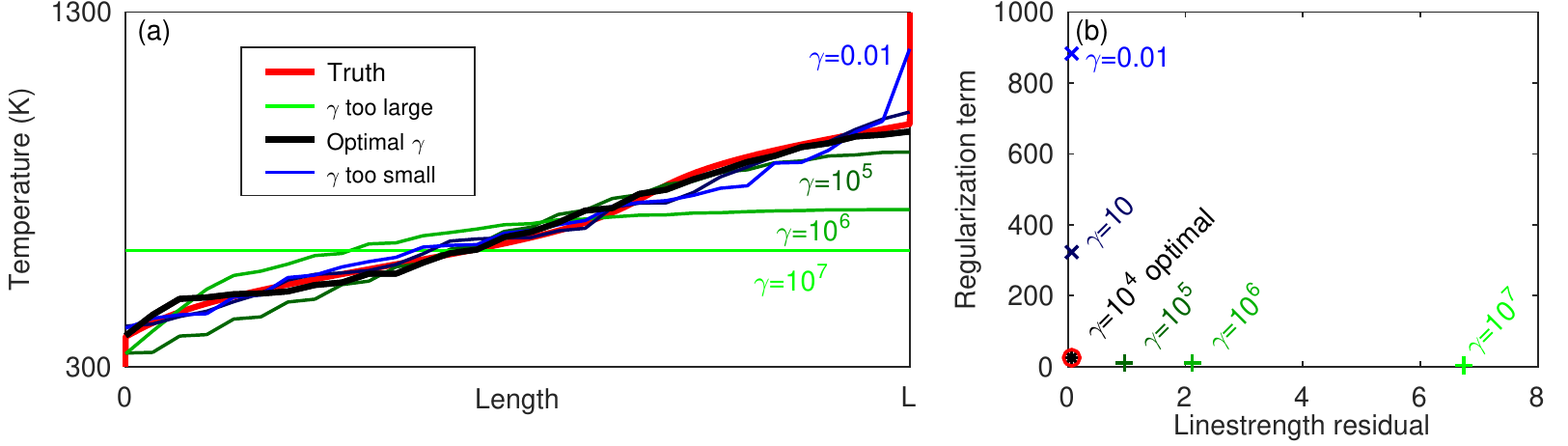}
\caption{Temperature solution depends on value of regularization parameter. (a) Temperature distribution retrievals for several regularization parameters. (b) L-curve of $2^{nd}$-order Tikhonov regularization term vs linestrength measurement term for each temperature distribution in (a). Small-$\gamma$ solutions overestimate maximum temperature, just like the TSVD solutions in Fig. \ref{fig:tsvd}. The most accurate temperature retrievals lie at the corner of the L-curve closest to the origin.}
\label{fig:lcurve}
\end{figure}

The challenge of Tikhonov regularization is selecting an optimal intermediate value of the regularization parameter $\gamma$ to retrieve the most accurate temperature distributions without knowing the true temperature distribution. Out of several $\gamma$-selection methods \protect\cite{tikhonovfirst, hansentext}, the L-curve method depicted in Fig. \ref{fig:lcurve} selects an accurate temperature CDF\textsuperscript{-1} for all four test cases. This L-curve method works as follows. First the $2^{nd}$-order length-binning inversion is calculated for several values of $\gamma$ spanning several orders of magnitude, using a simulated annealing algorithm based on Corana et al [46] to determine the minimum of Eq. \eqref{eq:tikh}. Then the two terms of the minimization function, measurement residual and temperature gradient, are plotted for each $\gamma$ on the two axes of the  L-curve in Fig. \ref{fig:lcurve}b. The corresponding temperature distributions are shown as monotonically-increasing discrete solutions in Fig. \ref{fig:lcurve}a. The small-$\gamma$ solutions appear in blue as chaotic and sharp-cornered in the temperature distribution panel, and as high-gradient, low-residual solutions on the L-curve panel. The large-$\gamma$ solutions appear in green as uniform and parabolic distributions on the temperature distribution panel, and as low-gradient, high-residual solutions on the L-curve panel. The best solution lies at the corner of the L-curve \protect\cites{invtext3, hansentext}, determined by a rectangular distance equation described in the appendix. This $\gamma$-selection is justified by Occam’s Razor, whereby of all the solutions with nearly equivalent fits to the linestrength data (those points which lie upon the vertical portion of Fig. \ref{fig:lcurve}b), the simplest solution with the smallest temperature gradients is the most likely. Indeed, even though the true temperature CDF\textsuperscript{-1} was not used in the L-curve inversion method, Fig. \ref{fig:lcurve} shows that the corner of the L-curve corresponds to the most accurate solution. In situations where the corner of the L-curve may be smoother and more ambiguous, one can select a range of plausible $\gamma$ and incorporate this spread in the temperature CDF\textsuperscript{-1} into the uncertainty bars of the inversion.

Appendix \ref{sec:TikhA} describes the regularization algorithm in more detail.

\subsection{Tikhonov regularization results and discussion}

\begin{figure}[h]
\includegraphics{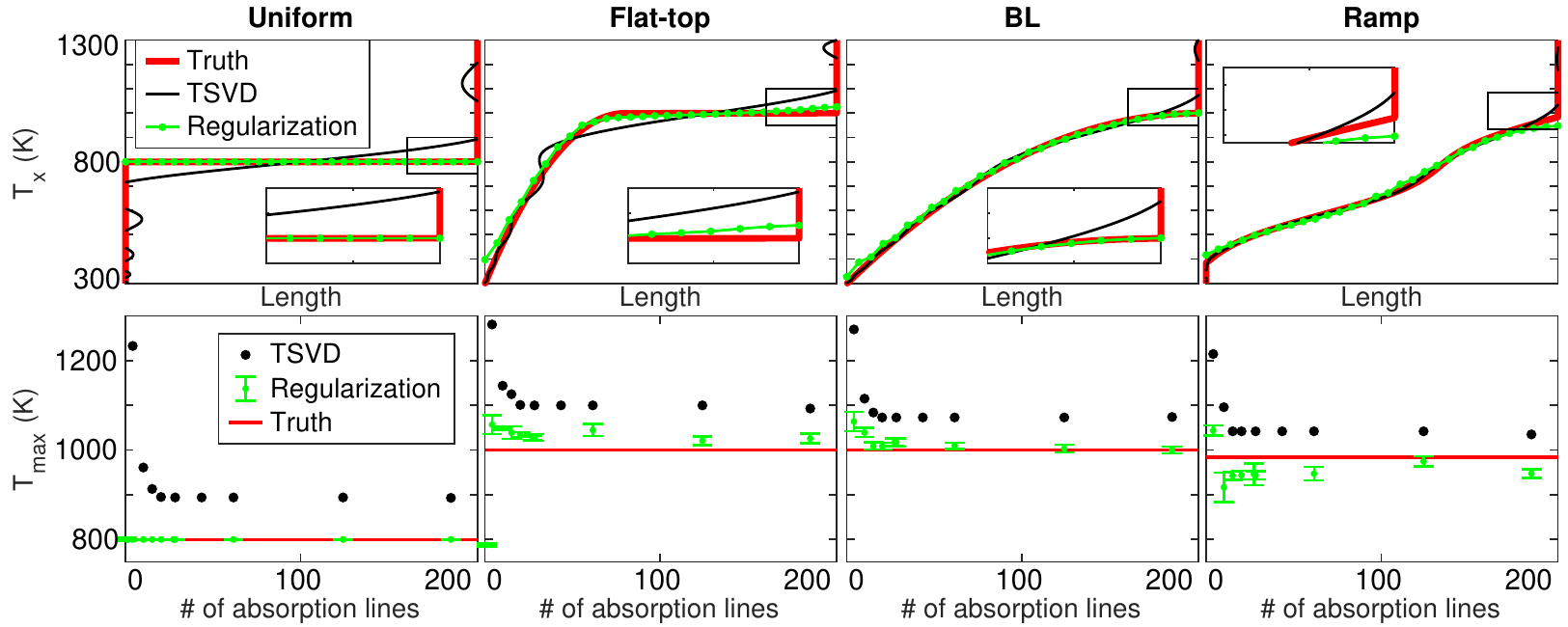}
\caption{(Top) Temperature distributions for different bandwidths determined using final inversion algorithm ($2^{nd}$-order Tikhonov regularization method with L-curve selection from Fig. \ref{fig:lcurve}). Regularization result (green) is shown as mean of 100 algorithm runs, each dot representing 1 length bin. (Bottom) Maximum temperature retrieved from above regularization method for 185 lines compared against TSVD method in Fig. \ref{fig:tsvd}. Inversion accuracy asymptotes by 20 absorption lines.}
\label{fig:results}
\end{figure}

Finally, we repeat the bandwidth study from Section \ref{sec:tsvd_results} using the L-curve regularization algorithm to determine improvements on the TSVD solution. The normalized linestrength inputs are simulated without noise for each bandwidth and for all four temperature distributions. The resulting Tikhonov regularization distributions are shown in Fig. \ref{fig:results}(top) for the 185-line inversion. The maximum temperature of the inversion is plotted in Fig. \ref{fig:results}(bottom) to assess this regularization method against the TSVD solutions as a function of the number of observable H\textsubscript{2}O absorption transitions in an expanding bandwidth around 6916 cm\textsuperscript{-1}.

The results in Fig. \ref{fig:results}(bottom) show the regularization algorithm improves the solution relative to the simpler TSVD inversion algorithm for all bandwidths and for all test cases. In particular, the regularization method found the correct uniform distribution where the truncated SVD solution produces a nonuniform-temperature solution. However, even the 185-line measurement with no error and Tikhonov regularization was not able to converge to the exact distribution for all test cases, tending to an intermediate parabolic curvature about $T_{max}$ which overestimates the flat-top distribution and underestimates the ramp distribution. The largest temperature deviations for Tikhonov regularization across these four test cases lie within 3\% of the true distribution.

\section{Summary}
Absorption spectra spanning many absorption transitions can determine the nonuniform temperature distribution along the laser path through an inversion process.
We show that the inversion process relates two continuous functions. The continuous spectroscopic function is the normalized linestrength of the absorption transitions as a function of their lower-state energy. The piecewise-continuous temperature distribution function is of the density of light-absorbing molecules at each temperature.

Previously it was believed that one could solve the temperature distribution more accurately as the spectrometer measured more transitions. In this paper we show that only the first $\sim$14 perfectly-measured transitions at different $E^{\prime\prime}$ provided additional information about the temperature distribution.
From singular value decomposition, the smoothly-varying normalized linestrength function can never resolve the discontinuous temperature distribution to a better approximation than 10-20 Fourier-series-like basis vectors.

However, as we show in the companion paper, a real broadband absorption measurement may benefit from utilizing more than 10-20 features to mitigate measurement uncertainty—by the standard error equation, additional uncertain measurements may reduce the uncertainty of the shape of the normalized linestrength function. If more spectral bandwidth reduces the uncertainty in the curvature of the normalized linestrength function, by sampling different lower-state energies or providing high-SNR absorption features, the temperature distribution may improve. The companion paper will also show how to extract the normalized linestrength curve from an absorption spectrum.

The combustion practitioner often wants $T_{max}$, the maximum temperature along the path. Unfortunately, we discovered that the temperature inversion from single-beam absorption spectroscopy is most uncertain at $T_{max}$. This issue occurs because the temperature distribution function must be discontinuous at $T_{max}$, and the SVD solution to the temperature inversion, which behaves similarly to a Fourier-series approximation, will be most inaccurate at discontinuities. Therefore, the temperature inversion will overshoot its $T_{max}$ estimation even as the inversion may resolve the rest of the temperature distribution away from $T_{max}$. To combat this tendency to overestimate $T_{max}$, we must apply constraints to our temperature inversion.

To improve $T_{max}$ estimation, we implement a Tikhonov regularization algorithm. This algorithm assumes the gas is uniform unless there is adequate spectroscopic evidence for temperature nonuniformity, and then will prefer certain temperature distribution shapes. This algorithm may not be appropriate for all applications, but it does work well for our four test cases. While computationally expensive, the Tikhonov algorithm returned four different temperature distributions to within 3\%  accuracy. 
\\
\bigskip
\\
\textit{Acknowledgments}: The authors wish to thank Ian Grooms for helpful mathematical discussion, Julie Steinbrenner for manuscript edits, and the reviewers for helpful suggestions. This work was funded by the National Science Foundation under grant CBET 1454496 and the Air Force Office of Scientific Research under grant FA9550-17-1-0224.

\appendix
\renewcommand{\theequation}{\thesection.\arabic{equation}}
\renewcommand{\thefigure}{\thesection.\arabic{figure}} 
\renewcommand{\thetable}{\thesection.\arabic{table}}
\setcounter{equation}{0}
\setcounter{figure}{0}
\section{Availability of material}
The figures in this article, as well as the data and MATLAB plotting scripts used to produce them, are available openly under the CC-BY license \protect\cite{data}. \protect\cite{data} also includes Lbin\_quick.m, our first MATLAB implementation of the Tikhonov regularization algorithm in Section \ref{sec:tikh}. This same function is available in Python (length\_bin.py) as part of our NTFit package \protect\cite{ntfit}. NTfit is open-source under the BSD-3 license, and will be expanded for broadband measurements with the companion paper.

\section{Calculating temperature distribution functions (TDFs)}
\label{sec:pdfexact}
Section \ref{sec:pdf} argued that the integrated area of an absorption feature could be calculated by a space-integral or a temperature-integral. This section shows how to perform the change-of-variables in Eq. \eqref{eq:integral} for all cases.
\begin{equation}
\label{eq:integral}
\int_\nu \alpha_i = \underbrace{\int_0^L P_{H_2O}(x) S_i(T(x)) dx}_\textrm{Eq. 3} = \underbrace{\overline{P_{H_2O}}L \int_0^\infty TDF(T) S_i(T) dT}_\textrm{Eq. 4}
\end{equation}

We move the total column density $\overline{P_{H_2O}}L$ out of the temperature integral, and define it by the relationship
\begin{equation}
\overline{P_{H_2O}}L = \int_0^L P_{H_2O}(x) dx
\end{equation}
After removing this normalization term out of the integral, we define the $TDF(T)$ as some fractional density of H\textsubscript{2}O molecules along the path at a particular temperature $T$. The column density $\overline{P_{H_2O}}L$ allows the TDF to have unity area. For nonuniform concentrations, where $P_{H_2O}$ = $f(x)$, the CDF is not a simple monotonic version of the temperature profile, but the fraction of H\textsubscript{2}O molecules below $T$, calculated from the TDF in Eq. \eqref{eq:toCDF}
\begin{equation}
\label{eq:toCDF}
CDF(T) = \int_0^T TDF(T) dT
\end{equation}

We can calculate the TDF from the $T(x)$ and $P_{H_2O}(x)$ profiles. When $T(x)$ has an inverse $x(T)$, then we can apply this inverse function in variable substitution of Eq. \eqref{eq:path}
\begin{equation}
\int_\nu \alpha_i = \int_{T(x=0)}^{T(x=L)} P_{H_2O}(x(T)) S(T) \frac {1} {x^\prime (T)} dT
\end{equation}
Setting this integral equal to Eq. \eqref{eq:inversiondT}, we get the following relationship for $TDF(T)$ from an inverse temperature profile $x(T)$:
\begin{equation}
\label{eq:pdfdef}
TDF(T) = \left( \frac {P_{H_2O}(x(T))} {\overline{P_{H_2O}}L \ \ x^\prime (T)} \right) 
\end{equation}

While $T(x)$ does not always have an inverse, it is piecewise-invertible in smaller monotonic segments. In this case, we split the integral \eqref{eq:pdfdef} into multiple pieces. Rather than integrate from $T(x=0)$ to $T(x=L)$, integrate from $T(x=0)$ to $T(x=L/2)$, or between all local extrema of the function $T(x)$. The $TDF(T)$ adds contributions from each invertible segment $x(T)$.

The other complication is uniform pieces of the function $T(x)$, for instance the central portion of the flat-top profile in Fig. \ref{fig:pdf}. We add the TDF contributions from these uniform segments with the delta function $\delta(T)$.
\begin{equation}
TDF(T) \mathrel{+}= \delta(T) \frac {\int_{x_1}^{x_2} P_{H_2O}(x) dx}{\overline{P_{H_2O}}L}
\end{equation}

The temperature profile $T(x)$ is not always known as a continuous function. Thermocouple measurements provide a discrete set of $N$ temperatures, from which one might fit a function.
The Tikhonov regularization in Section \ref{sec:tikh} produces an array $T(x_i)$, and the tomography in \protect\cite{caima} additionally produces $P_{H_2O}(x_i)$.
In this case, we sum the $N$ contributions to the TDF in Eq \eqref{eq:tikhpdf}.

\begin{equation}
\label{eq:tikhpdf}
TDF(T_j) = \sum_{i \in I} \left( \frac{P_{H_2O}(x_i)}{\overline{P_{H_2O}}L} \Delta x_i \right) \ , \  \ I = \{1 \le i \le N \ | \ T(x_i) = T_j \}
\end{equation}

This produces a TDF much like the the non-negative least-squares result in Fig. \ref{fig:lsq}b, but if $N$ is large enough then the CDF\textsuperscript{-1} appears smooth.
The Tikhonov regularization solutions in Figs. \ref{fig:lcurve}, \ref{fig:results} have $N$ = 30, which appears adequate for a smooth CDF\textsuperscript{-1}. The Tikhonov solutions in Section \ref{sec:tikh} also satisify the unity area requirement of distribution functions $\int_0^\infty TDF dT = 1$.
Section \ref{sec:tikh} solutions have $P_{H_2O}(x_i) = \overline{P_{H_2O}}$, so the TDFs have $N$ = 30 positive chunks of height $1/N$ and width 1 Kelvin.

\setcounter{equation}{0}
\setcounter{figure}{0}
\section{Line-of-Sight average temperature bias in absorption spectroscopy due to path nonuniformity}
\label{sec:2line}
Absorption spectroscopy, including traditional two-line thermometry, normally aims to measure a single path-average temperature along the laser line-of-sight. Using our normalized linestrength concept, we will discuss the temperature bias that nonuniformity imposes on traditional absorption spectroscopy mean-temperature measurements. 

\subsection{Two-line thermometry bias}
Two-line thermometry retrieves the path-average temperature at which both normalized linestrengths are equal, as formulated in Eq. \eqref{eq:2line}. Eq. \eqref{eq:2line} follows from Eq. \eqref{eq:path} for a uniform path when $T(x) = \overline{T}$.
 \begin{equation}
\label{eq:2line}
\frac {\int_\nu \alpha_1} {S_1(\overline{T})} = \frac {\int_\nu \alpha_2} {S_2(\overline{T})} = \overline{P\chi} L 
\end{equation}

However, this spectroscopy-weighted mean temperature may differ from the true path-averaged temperature for nonuniform measurements.
We use the normalized linestrength curves to explain this mean-temperature bias that can arise from Eq. \eqref{eq:2line}.

\begin{figure}
\centering
\includegraphics{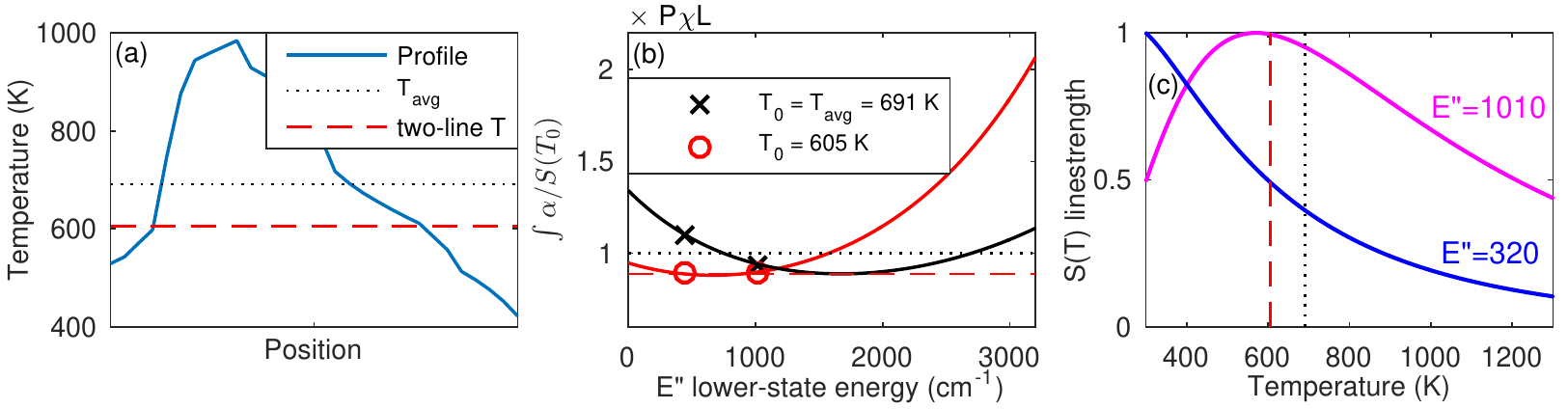}
\caption{Two-line thermometry imposing -86 Kelvin bias on ramp profile (a) Ramp temperature profile (b) Normalized linestrength curves for same ramp profile evaluated at two relevant temperatures (c) Linestrength temperature-dependence for two $E^{\prime\prime}$ used for denominator of x-marked points in (b). See supplemental video, SE\_nonuni.avi, that animates the normalized linestrength curve for both a nonuniform (solid red traces) and a path-average (solid black traces) temperature profile at many more reference temperatures $T_0$ (dotted traces).}
\label{fig:2line}
\end{figure}

One can visualize two-line thermometry by adjusting the reference temperature $T_0$ of the normalized linestrength curve, shown in the Fig. \ref{fig:2line} video.
Of the components of normalized linestrength $\hat{S} = \int \alpha / S(T_0)$, the temperature distribution sets the integrated areas $\int \alpha$, but the reference linestrength $S(T_0)$ changes with $T_0$.
Fig. \ref{fig:2line} highlights two normalized linestrengths on Fig. \ref{fig:2line}b, with their corresponding reference linestrengths $S(T_0)$ shown in Fig. \ref{fig:2line}c.
As $T_0$ increases, $S(T_0)$ in Fig. \ref{fig:2line}c decreases more rapidly for the lower-$E^{\prime\prime}$ feature.
$S(T_0)$ is the denominator of the normalized linestrengths $\hat{S}$ in Fig. \ref{fig:2line}b, so $\hat{S}$ increases more rapidly for the lower-$E^{\prime\prime}$ feature, and the slope of the normalized linestrength curve $d\hat{S}/dE^{\prime\prime}$ decreases.

To satisfy the two-line thermometry condition (Eq. \eqref{eq:2line}) on the normalized linestrength plot, the normalized linestrength at the two measured lower-state energies must have the same y-axis value.
This technique assumes that the entire normalized linestrength curve becomes horizontal where $T_0 = \overline{T}$.
However, a nonuniform temperature distribution will produce a curved normalized linestrength function even when $T_0 = \overline{T}$ (see Fig. \ref{fig:snorm}e).
So depending on the two selected $E^{\prime\prime}$ points, the $T_0$ which satisfies Eq. \eqref{eq:2line} might not correspond to the true path-average temperature $\overline{T}$.

We demonstrate this temperature bias in Fig. \ref{fig:2line} for the temperature profile in Fig. \ref{fig:2line}a.
Two-line thermometry only measures two points of the normalized linestrength curve (the x-marks in Fig. \ref{fig:2line}b), but we overlay the entire curve to demonstrate temperature bias.
A two-line selection of $E^{\prime\prime}$ = 315 and 1006 shifts the reference temperature by -86 Kelvin relative to the true mean temperature in order to satisfy Eq. \eqref{eq:2line}.
This reduction of $T_0 = \overline{T}-86$ flattens the normalized linestrength curve in Fig. \ref{fig:2line}b for $E^{\prime\prime}<1000$, but makes the curve much steeper for $E^{\prime\prime}>1000$.
In other words, Eq. \eqref{eq:2line} mistakenly selects a $T_0$ which flattens the sampled region of Fig. \ref{fig:2line}b at the expense of increased slope of the overall curve.
For this temperature distribution, a better two-line selection is $E^{\prime\prime}$ = 880 and 3060, where the linestrength curve normalized by the correct $\overline{T}$ crosses $P \chi L$.

\subsection{Broadband thermometry bias}
Two-line thermometry often mistakes the path-average gas temperature because it only measures two points of the normalized linestrength curve. Following this reasoning, we might expect a broadband measurement to produce more reliable temperatures, as broadband spectra measure many more points across the normalized linestrength curve. If we use more complicated temperature inversions, this is indeed the case. However, traditional broadband-fitting methods will tend to underestimate the path-average temperature of a nonuniform path. Here, we derive how traditional broadband thermometry appears in the new framework of the normalized linestrength curve. This explains how temperature nonuniformity will bias those spectral-processing techniques that do not account for path-nonuniformity.

Traditionally, one determines temperature from one spectral-fitting step, by adjusting the temperature and concentration of an absorption model to best fit the measured absorption spectrum (Eq. \eqref{eq:spectrum}).
\begin{equation}
\label{eq:spectrum}
\min_{T,P\chi L} \left\{ \sum_{i=1}^{frequencies} \left(\alpha_{meas} - \alpha_{model} \right)^2 \right\}
\end{equation}

For a precise fit, the model term $\alpha_{model}$ comes from Eq. \eqref{eq:beer}. In order to demonstrate the mean-temperature bias, we will ignore lineshape effects and approximate Eq. \eqref{eq:spectrum} as a fit to each of the integrated areas in the spectrum.
\begin{equation}
\min_{T,P\chi L} \left\{ \sum_{i=1}^{lines} \left(\int\alpha_{meas} - S_i P \chi L \right)^2 \right\}
\end{equation}

We then factor out the linestrength term to make the normalized linestrength appear explicitly in Eq. \eqref{eq:broadband}.
\begin{equation}
\label{eq:broadband}
\min_{T,P\chi L} \left\{ \sum_{i=1}^{lines} S_i^2(T) \left(\frac{\int\alpha_{meas}}{S_i(T)} - P\chi L \right)^2 \right\}
\end{equation}
The terms in the parentheses of Eq. \eqref{eq:broadband} appear like the two-line thermometry relationship (Eq. \eqref{eq:2line}). Eq. \eqref{eq:broadband} simplifies to zero when the two-line thermometry condition (Eq. \eqref{eq:2line}) is satisfied, when each measured point on the normalized linestrength curve $\int \alpha / S_i$ is equal to the same $P\chi L$. When this occurs, the normalized linestrength curve is horizontal. However, we know that for nonuniform-temperature paths, the normalized linestrength curve cannot be perfectly flat, so Eq. \eqref{eq:broadband} must be greater than zero for all $T, P\chi L$. In this instance, the non-zero minimum of Eq. \eqref{eq:broadband} will not generally correspond to the path-average temperature.

There are three components to Eq. \eqref{eq:broadband}. The integrated area term is the normalized linestrength, representing one point on the curve in Fig. \ref{fig:2line}b. The $P\chi L$ column density represents the y-axis value of the dashed horizontal line on the normalized linestrength curve \ref{fig:2line}b. The least-squares fit is adjusting the reference temperature $T_0$ to squash the normalized linestrength curve as flat and horizontal as possible. Finally, the linestrength $S(T)$ factored outside of the parentheses is a weight, ensuring that the stronger absorption features have more influence upon the temperature fit. A broadband spectral fit acts like a least-squares fit to each of the integrated areas of the individual spectral features, where the stronger absorption features will disproportionately influence the fit.  

In most spectra, the stronger absorption features have lower $E^{\prime\prime}$. Thus Eq. \eqref{eq:broadband} will tend to flatten the low-$E^{\prime\prime}$ portion of the curve at the expense of an increase in slope of the lower-weighted data points at the high-$E^{\prime\prime}$ part of the curve. Even if the broadband spectrum measures the entire normalized linestrength curve (Fig. \ref{fig:2line}b), the single-temperature spectral fit will underestimate the path-average temperature due to the unequal weights $S_i(T)$.

If you have some estimate of the temperature nonuniformity, you can select the spectral bandwidth to minimize this temperature bias. We include a supplemental Python code, $temperature\_bias$, to estimate this mean-temperature bias for a particular bandwidth of H\textsubscript{2}O absorption features. This code uses Eq. \eqref{eq:broadband} with some HITRAN linelist selection (using the HAPI package \protect\cite{hapi}) of  lower-state energies and reference linestrengths to estimate mean-temperature bias.

\setcounter{equation}{0}
\setcounter{figure}{0}
\setcounter{table}{0}
\section{Regularization algorithm}
\label{sec:TikhA}
The matrix inversion algorithm to extract a temperature distribution from a set of normalized linestrengths, which was introduced in Section \ref{sec:tikh}, is described here in additional detail. Briefly, a nonlinear random-search minimization algorithm \protect\cite{simanneal} is performed for a variety of regularization parameters, and then the best solution is determined to lie at the corner of the L-curve (Fig. \ref{fig:lcurve}) in order to produce a temperature distribution which both maintains plausibly small temperature gradients and fits the absorption data. Our algorithm and this appendix follow the flowchart in Table \ref{tab:tikh}. \\

\begin{table}[h]
\begin{small}
\begin{tabular}{p{0.5mm} p{4.6cm} p{3.5cm} p{6.5cm}}
 \toprule
& Algorithm step description      &  Computation method   & Parameter outputs \\
 \midrule 
1) & \makecell[tl]{Calculate regularization \\ limiting cases} & Least-squares & 
        \makecell[tl]{Least-squares residual $res_{LSQ}$ \\ and gradient $reg_{LSQ}$ \\  Uniform-temperature residual $res_u$ \\ and gradient $0$} \\
 \addlinespace[0.2cm]

2) & \makecell[tl]{Set up intermediate \\ regularization cases} & N/A     & 
         \makecell[tl]{Regularization parameters $\gamma$ \\ Annealing temperature $T_{SA}$ \\Initial temperature profile $T(x)_0$}\\ 
 \addlinespace[0.2cm]

3)& \makecell[tl]{Perform inversion for each \\ regularization parameter $\gamma$} & 
       \makecell[tl]{Simulated annealing \\ to minimize Eq. \ref{eq:tikh_full}}
    & \makecell[tl]{Output temperature profile $T(x)_\gamma$ \\ Linestrength residual $res_\gamma$\\  Profile gradient $reg_\gamma$ \\}\\
 \addlinespace[0.2cm]

4)& \makecell[tl]{Determine best inversion \\ and other plausible solutions} & L-curve Eq. \ref{eq:corner} & accepted distributions with $\gamma_{best}$ \\
 \bottomrule
\end{tabular}
\end{small}
\caption{Flowchart for Tikhonov regularization algorithm used for inversions in Figs. \ref{fig:lcurve} and \ref{fig:results}}
\label{tab:tikh}
\end{table}

1) \underline{Calculate limiting cases}\\

First, the two limiting cases for length-binning regularization are calculated.  These limiting cases are uniform-temperature and non-negative least-squares. This step is much faster than the inversions in step 3, and helps to interpret the inversion results. The best uniform-temperature retrieval is determined using an exhaustive domain search. The least-squares solution is found using MATLAB’s $lsqnonneg()$ algorithm \protect\cite{lsq}, which outputs the H\textsubscript{2}O column density TDF at each temperature from 300-1300 Kelvin. If the least-squares solution does not fit the linestrength measurement to at least a 7\% reduction in the 2-norm residual relative to the uniform solution, we decide that absorption spectroscopy provides insufficient evidence for temperature nonuniformity, and we select the uniform-temperature solution without performing the Tikhonov regularization algorithm. This 7\% guideline is a result of the tendency, as predicted by TSVD in Fig. \ref{fig:tsvd}a, of truly uniform profiles to produce nonuniform retrievals from absorption spectroscopy measurements with insufficient number of absorption features or SNR.\\

2) \underline{Set up intermediate regularization cases}\\

When the limiting cases demonstrate sufficient evidence for temperature nonuniformity, we continue with an iterative regularization algorithm. Our Tikhonov regularization algorithm uses a random-search method rather than a direct matrix inversion. This length-binning method determines the combination of 30 different temperatures with the best fit to the minimization function (Eq. \eqref{eq:tikh_full}) by an improved trial-and-error method known as simulated annealing. 
\begin{equation}
\label{eq:tikh_full}
\min_{P\chi L, T(x_b), \gamma} \left\{\left\lVert \frac{\int_\nu \alpha_i}{S_i(T_0)(P \chi L)_0} - \frac{\overline{P \chi}L_{act}}{P_0 \chi_0 L_0} \sum\limits_{b=1}^N \frac{S_i (T(x_b))}{S_i(T_0)}  \right\rVert + \gamma \lVert \mbox{temperature gradient} \rVert  \right\}
\end{equation}

The solution space contains $1001^{30}$ solutions, as each length-bin $T(x_b)$ can lie at one of 1001 different temperatures (from 300-1300 Kelvin), and the arrangement of those temperatures in space influences the temperature gradient regularization term.
In addition to this discrete temperature distribution, a H\textsubscript{2}O column density scaling factor, $pxl$, floats to allow for an unknown uniform column density $(\overline{P \chi}L)_{act} / (P \chi L)_0$.

The regularization algorithm in step 3 requires three inputs. The first input, regularization parameters $\gamma$, should vary over several orders of magnitude. We typically vary over 5 orders of magnitude, and later add sub-order-of-magnitude refinement around the optimal regularization corner value. The second input, the initial temperature profile $T(x)$, does in practice influence the result, because the solution space is so large that the algorithm typically finds a local rather than global minimum. A logical initial-profile choice is either the least-squares or uniform solution from step 1. We typically initialize with the uniform profile, because it is more likely to find a non-uniform solution than the least-squares initialization is to converge to the uniform solution. The last input, the simulated annealing temperature $T_{SA}$, must be large enough that the algorithm does not automatically return the initial temperature profile. To test whether this parameter is too small, initialize with the least-squares temperature profile and a large enough regularization parameter that the minimization function should return the uniform profile. If the annealing temperature is large enough, then the output temperature profile will be very different from the least-squares initial value.\\

3) \underline{Inversion at each regularization value}\\

The simulated annealing search algorithm \protect\cite{simanneal} must constrain the maximum temperature to occur in an intermediate length-bin for our purposes. Otherwise, the 2\textsuperscript{nd}-order regularization forces a ramp rather than a uniform profile. We find that the search conditions are adequate to enforce this condition, without using an explicit constraint.  Our search condition imposes a +5 Kelvin bias on the central length-bin test point, and a negative bias on the first and final length bins, in addition to the Metropolis-Hastings Gaussian search used in Corana et al \protect\cite{simanneal}. This bias in the search temperature is more important for single-beam measurements than for tomographic reconstruction over multiple beam paths. If a tomography measurement contains beam paths that skirt along the colder edges in addition to paths that cut across the hottest regions, then the solution will naturally be hottest in the middle of the two-dimensional area.

This simulated annealing algorithm takes $\sim 1$ minute for each minimization, and the minimizations for different regularization parameters can be performed on parallel processers. Because this search space is so large, with a nonlinear relationship on the minimization function, the minimization algorithm does not always locate the true best-fit temperature distribution. Instead there remains some scatter in the minimization output for subsequent simulated annealing minimizations of the same data with the same regularization parameter. This scatter in the minimization algorithm output is responsible for the uncertainty bars in Fig. \ref{fig:results}, and also makes traditional finite-difference maximum-curvature schemes for regularization parameter selection \protect\cite{hansentext} infeasible.

In length-binning, regularization is acting on a temperature profile; however, only the regularization term, not the spectroscopy model, is sensitive to the permutation of temperatures across the different length-bins. Thus the location of $T_{max}$ in the length-binning output is degenerate to the linestrength fit, \textit{so we rearrange this length-binning output into a monotonically increasing discrete CDF\textsuperscript{-1}.} \\

4) \underline{Determine best inversion}\\

Our regularization parameter selection is determined by a Euclidean distance method defined in Eq. \eqref{eq:corner}. This equation can be interpreted as a normalized distance of the inversion result from the corner of a rectangle defined by the limit-case inversions in step 1. A true L-shaped curve contains a corner at right angle to the two limiting cases, so the temperature gradients ($\lVert$ reg $\rVert$) will be the same as the uniform case, and the linestrength measurement residual ($\lVert$ res $\rVert$) will match the least-squares case. The difference in linestrength residual and regularization from this theoretical corner is normalized by the limiting cases, so that the least-squares solution and the uniform solution ($\gamma \to 0$ and $\gamma \to \infty$) are both a distance 1 from this corner. The best solution is at some intermediate regularization parameter which lies at the minimum of Eq. \eqref{eq:corner}. 

\begin{equation}
\label{eq:corner}
\gamma_{corner} = \min_\gamma \left\{ \left( \frac {\lVert \mathrm{res} \rVert_\gamma - \lVert \mathrm{res} \rVert_{LSQ}} {\lVert \mathrm{res} \rVert_U}  \right)^2 + \left( \frac {\lVert \mathrm{reg} \rVert_\gamma - 0} {\lVert \mathrm{reg} \rVert_{LSQ}}  \right)^2 \right\}
\end{equation}
Real L-curves will be more rounded than L-shaped, particularly for noisy measurements. This corner algorithm allows a guideline for uncertainty in the regularization parameter selection. Because of the scatter in the solution from an imperfect simulated annealing algorithm, many different solutions at different regularization parameters may each lie close to the corner of the L-curve. As a guideline, every solution with an Eq. \eqref{eq:corner} solution within 0.07  of the best-fit regularization solution can be interpreted as an equally plausible temperature distribution. Then the scatter in the temperature distributions of each of those good-fit solutions (in the set $\gamma \in \gamma_{best}$) can form the uncertainty bars in the retrieved temperature distribution. 


\begin{thebibliography}{10}
\bibitem{aero}	D. Zhao, Z. Lu, H. Zhao, X. Y. Li, B. Wang, and P. Lu, “A review of active control approaches in stabilizing combustion systems in aerospace industry,” Prog. Aerosp. Sci., vol. 97, pp. 35–60, 2018, doi: 10.1016/j.paerosci.2018.01.002.
\bibitem{hotstreak1}	T. Povey and I. Qureshi, “Developments in Hot-Streak Simulators for Turbine Testing,” J. Turbomach., vol. 131, no. 3, Jul. 2009, doi: 10.1115/1.2987240.
\bibitem{turbine1}	M. D. Barringer, K. A. Thole, and M. D. Polanka, “Experimental evaluation of an inlet profile generator for high-pressure turbine tests,” J. Turbomach., vol. 129, no. 2, pp. 382–393, 2007. doi: 10.1115/1.2436897
\bibitem{turbine2}	C. M. Cha, S. Hong, P. T. Ireland, P. Denman, and V. Savarianandam, “Experimental and Numerical Investigation of Combustor-Turbine Interaction Using an Isothermal, Nonreacting Tracer,” J. Eng. Gas Turbines Power, vol. 134, no. 8, Aug. 2012, doi: 10.1115/1.4005815.
\bibitem{hotstreak}	A. Andreini, T. Bacci, M. Insinna, L. Mazzei, and S. Salvadori, “Hybrid RANS-LES modeling of the aerothermal field in an annular hot streak generator for the study of combustor–turbine interaction,” J. Eng. Gas Turbines Power, vol. 139, no. 2, 2017. doi: 10.1115/1.4034358
\bibitem{doll}	U. Doll, M. Dues, T. Bacci, A. Picchi, G. Stockhausen, and C. Willert, “Aero-thermal flow characterization downstream of an NGV cascade by five-hole probe and filtered Rayleigh scattering measurements,” Exp. Fluids, vol. 59, no. 10, p. 150, Sep. 2018, doi: 10.1007/s00348-018-2607-z.
\bibitem{flametreatment}	S. Farris, S. Pozzoli, P. Biagioni, L. Duo, S. Mancinelli, and L. Piergiovanni, “The fundamentals of flame treatment for the surface activation of polyolefin polymers - A review,” Polymer, vol. 51, pp. 3591–3605, 2010. doi: 10.1016/j.polymer.2010.05.036
\bibitem{palaghita}	T. I. Palaghita and J. M. Seitzmann, “Control of Temperature Nonuniformity Based on Line-of-Sight Absorption,” in AIAA Joint Propulsion Conference, Ft. Lauderdale, FL, Jul. 2004, vol. 40. doi: 10.2514/6.2004-4163
\bibitem{goldensteinreview}	C. S. Goldenstein, R. M. Spearrin, J. B. Jeffries, and R. K. Hanson, “Infrared laser-absorption sensing for combustion gases,” Prog. Energy Combust. Sci., vol. 60, pp. 132–176, 2016. doi: 10.1016/j.pecs.2016.12.002
\bibitem{2line}	R. K. Hanson and P. K. Falcone, “Temperature measurement technique for high-temperature gases using a tunable diode laser,” Appl. Opt., vol. 17, no. 16, pp. 2477–2450, Aug. 1978. doi: 10.1364/AO.17.002477
\bibitem{goldensteinnonuni}	C. S. Goldenstein, I. A. Schultz, J. B. Jeffries, and R. K. Hanson, “Two-color absorption spectroscopy strategy for measuring the column density and path average temperature of the absorbing species in nonuniform gases,” Appl. Opt., vol. 52, no. 33, pp. 7960–7972, Nov. 2013. doi: 10.1364/AO.52.007950
\bibitem{12}	X. Zhou, J. B. Jeffries, and R. K. Hanson, “Development of a fast temperature sensor for combustion gases using a single tunable diode laser,” Appl. Phys. B, vol. 81, no. 5, pp. 711–722, Sep. 2005. doi: 10.1007/s00340-005-1934-y
\bibitem{werblinski}	T. Werblinski, P. Fendt, L. Zigan, and S. Will, “High-speed combustion diagnostics in a rapid compression machine by broadband supercontinuum absorption spectroscopy,” Appl. Opt., vol. 56, no. 15, pp. 4443–4453, May 2017. doi: 10.1364/AO.56.004443
\bibitem{blfirst}	X. Ouyang and P. L. Varghese, “Line-of-sight absorption measurements of high temperature gases with thermal and concentration boundary layers,” Appl. Opt., vol. 28, no. 18, pp. 3979–3984, 1989. doi: 10.1364/AO.28.003979
\bibitem{15}	Lohden, B., Kuznetsova, S., Sengstock, K. et al., “Fiber laser intracavity absorption spectroscopy for in situ multicomponent gas analysis in the atmosphere and combustion environments,” Appl. Phys. B, vol. 102, pp. 331–344, 2011. doi: 10.1007/s00340-010-3995-9
\bibitem{wms}G. B. Rieker, J. B. Jeffries, and R. K. Hanson, “Calibration-free wavelength-modulation spectroscopy for measurements of gas temperature and concentration in harsh environments,” Appl. Opt., vol. 48, no. 29, pp. 5546–5560, 2009. doi: 10.1364/AO.48.005546
\bibitem{Sanders} Scott Sanders, J. Wang, J. B. Jeffries, and R. K. Hanson, “Diode-laser absorption sensor for line-of-sight gas temperature distributions,” Appl. Opt., vol. 40, no. 24, pp. 4404–4415, Aug. 2001. doi: 10.1364/AO.40.004404
\bibitem{lbin}	X. Liu, J. B. Jeffries, and R. K. Hanson, “Measurement of Nonuniform Temperature Distributions Using Line-of-Sight Absorption Spectroscopy,” AIAA J., vol. 45, no. 2, pp. 411–419, Feb. 2007, doi: 10.2514/1.26708.
\bibitem{kaminski}	J. Hult, R. S. Watt, and C. F. Kaminski, “High bandwidth absorption spectroscopy with a dispersed supercontinuum source.,” Opt. Express, vol. 15, no. 18, pp. 11385–11395, 2007. doi: 10.1364/OE.15.011385
\bibitem{vcsel}	Sanders, Scott T., “Wavelength-agile fiber laser using group-velocity dispersion of pulsed super-continua and application to broadband absorption spectroscopy,” Appl. Phys. B, vol. 75, pp. 799–802, 2002. doi: 10.1007/s00340-002-1044-z
\bibitem{wagner}	S. Burkle, A. Dreizler, V. Ebert, and S. Wagner, “Experimental comparison of a 2D laminar diffusion flame under oxy-fuel and air atmosphere,” Fuel, vol. 212, pp. 302–308, Jan. 2018. doi: 10.1016/j.fuel.2017.10.067
\bibitem{tomoafrl}	L. Ma et al., “Tomographic imaging of temperature and chemical species based on hyperspectral absorption spectroscopy,” Opt. Express, vol. 17, no. 10, pp. 8602–8613, May 2009. doi: 10.1364/OE.17.008602
\bibitem{beijing}	C. Liu, L. Xu, and Z. Cao, “Measurement of nonuniform temperature and concentration distributions by combining line-of-sight tunable diode laser absorption spectroscopy with regularization methods.,” Appl. Opt., vol. 52, no. 20, pp. 4827–4842, Jul. 2013. doi: 10.1364/AO.52.004827
\bibitem{hayden}	T. R. S. Hayden and Rieker, Gregory B, “Large amplitude wavelength modulation spectroscopy for sensitive measurements of broad absorbers,” Opt. Express, vol. 24, no. 24, pp. 27910–27921, Nov. 2016. doi: 10.1364/OE.24.027910
\bibitem{vcselafrl}	K. D. Rein, S. Roy, S. T. Sanders, A. W. Caswell, F. R. Schauer, and J. R. Gord, “Measurements of gas temperatures at 100 kHz within the annulus of a rotating detonation engine,” Appl. Phys. B, vol. 123, no. 3, p. 88, Mar. 2017, doi: 10.1007/s00340-017-6647-5.
\bibitem{sanders2}	S. T. Melin and S. T. Sanders, ``H\textsubscript{2}O absorption thermometry accuracy in the 7321-7596 cm\textsuperscript{-1} range studied in a heated static cell at temperatures up to 1723K," J. Quant. Spectrosc. Radiat. Transf., vol. 214, pp 1-7, Jul. 2018, doi: 10.1016/j.jqsrt.2018.04.012.
\bibitem{fielddcs}	P. J. Schroeder et al., “Dual Frequency Comb Laser Absorption Spectroscopy in a 16 MW Gas Turbine Exhaust,” Proc. Combust. Inst., vol. 36, Jun. 2016. doi: 10.1016/j.proci.2016.06.032
\bibitem{highspeed}	A. D. Draper et al., “Broadband dual-frequency comb spectroscopy in a rapid compression machine,” Opt. Express, vol. 27, no. 8, pp. 10814–10825, 2019. doi: 10.1364/OE.27.010814
\bibitem{gasifier}	P. Schroeder, A.S. Makowiecki and M.A. Kelley et al., “Temperature and concentration measurements in a high-pressure gasifier enabled by cepstral analysis of dual frequency comb spectroscopy,” Proc. Combust. Inst., 2020. doi: 10.1016/j.proci.2020.06.011
\bibitem{midirdcs}	A.S. Makowiecki, D.I. Herman, N. Hoghooghi et al., “Mid-Infrared Dual Frequency Comb Spectroscopy for Combustion Analysis from 2.8 to 5 Microns,” Proc. Combust. Inst., 2020. doi: 10.1016/j.proci.2020.06.195
\bibitem{foltydb}	L. Rutkowski et al., “An experimental water line list at 1950 K in the 6250-6670 cm-1 region,” J. Quant. Spectrosc. Radiat. Transf., vol. 205, pp. 213–219, Jan. 2018, doi: 10.1016/j.jqsrt.2017.10.016.
\bibitem{folty_vernier} C. Lu, C. Lu, F. S. Vieira, F. S. Vieira, F. M. Schmidt, and A. Foltynowicz, “Time-resolved continuous-filtering Vernier spectroscopy of H2O and OH radical in a flame,” Opt. Express, vol. 27, no. 21, pp. 29521–29533, Oct. 2019, doi: 10.1364/OE.27.029521.
\bibitem{exoplanet}	H. A. Knutson, D. Charbonneau, L. E. Allen, A. Burrows, and S. T. Megeath, “The 3.6-8.0 $\mu m$ Broadband Emission Spectrum of HD 209458b: Evidence for an Atmospheric Temperature Inversion,” Astrophys. J., vol. 673, no. 1, p. 526, 2008, doi: 10.1086/523894.
\bibitem{ren}	T. Ren and M. F. Modest, “Temperature Profile Inversion from Carbon-Dioxide Spectral Intensities Through Tikhonov Regularization,” J. Thermophys. Heat Transf., vol. 30, no. 1, pp. 211–218, 2016, doi: 10.2514/1.T4561.
\bibitem{cai_review}	W. Cai and C. F. Kaminski, “Tomographic absorption spectroscopy for the study of gas dynamics and reactive flows,” Prog. Energy Combust. Sci., vol. 59, pp. 1–31, Mar. 2017, doi: 10.1016/j.pecs.2016.11.002.
\bibitem{grauer}	S. J. Grauer, J. Emmert, S. T. Sanders, S. Wagner, and K. J. Daun, “Multiparameter gas sensing with linear hyperspectral absorption tomography,” Meas. Sci. Technol., vol. 30, no. 10, p. 105401, Aug. 2019, doi: 10.1088/1361-6501/ab274b.
\bibitem{hitran}	I. E. Gordon et al., “The HITRAN2016 molecular spectroscopic database,” J. Quant. Spectrosc. Radiat. Transf., vol. 203, pp. 3–69, Dec. 2017, doi: 10.1016/j.jqsrt.2017.06.038.
\bibitem{invtext1}	R. C. Aster, B. Borchers, and Thurber, Clifford H, “Why Inverse Problems Are Difficult,” in Parameter Estimation and Inverse Problems, Waltham MA: Elsevier, 2013, pp. 14, 19–22.
\bibitem{boltzplot1} M. Brown, D. Barone, W. Terry, T. Barhorst, and S. Williams, "Accuracy, precision, and scatter in TDLAS measurements," in \textit{48th AIAA Aerospace Sciences Meeting Including the New Horizons Forum and Aerospace Exposition}, p. 302, 2010. doi: 10.2514/6.2010-302
\bibitem{boltzplot2} J. France, "High temperature and pressure measurements from TDLAS through the application of 2nd derivative fitting and the aggregate Boltzmann method." PhD dissertation, U. of Michigan 2019. p. 92-94
\bibitem{invtext2}	R. C. Aster, B. Borchers, and C. H. Thurber, “Singular Value Decomposition,” in Parameter Estimation and Inverse Problems, Waltham MA: Elsevier, 2013, pp. 55-61,64-67, 81–87.
\bibitem{caima}	W. Cai, D. J. Ewing, and L. Ma, “Application of simulated annealing for multispectral tomography,” Comput. Phys. Commun., vol. 179, pp. 250–255, 2008. doi: 10.1016/j.cpc.2008.02.012
\bibitem{lsq}	C. L. Lawson and R. J. Hanson, “Solving Least Squares Problems,” in Solving Least Squares Problems, Englewood Cliffs, N.J.: Prentice-Hall, 1974, pp. 160–164.
\bibitem{invtext3}	R. C. Aster, B. Borchers, and Thurber, Clifford H, “Higher-order Tikhonov Regularization,” in Parameter Estimation and Inverse Problems, Waltham MA: Elsevier, 2013, pp. 93–95, 103–111.
\bibitem{tikhonovfirst}	L. Ma and W. Cai, “Determination of the optimal regularization parameters in hyperspectral tomography,” Appl. Opt., vol. 47, no. 23, pp. 4186–4192, Aug. 2008. doi: 10.1364/AO.47.004186
\bibitem{simanneal}	A. Corana, M. Marchesi, C. Martini, and S. Ridella, “Minimizing multimodal functions of continuous variables with the ‘Simulated Annealing’ algorithm,” ACM Trans. Math. Softw., vol. 13, no. 3, pp. 262–280, Sep. 1987, doi: 10.1145/29380.29864.
\bibitem{hansentext}	P. C. Hansen, Discrete Inverse Problems: Insight and Algorithms. SIAM, 2010.
\bibitem{data} [dataset] N.A. Malarich and G.B. Rieker. Figures, plotting scripts, and data for "Resolving non-uniform temperature distributions with single-beam absorption spectroscopy. Part I: Theoretical capabilities and limitations" Zenodo, 2020. doi: 10.5281/zenodo.3939090
\bibitem{ntfit} [dataset] Nonuniform-Temperature Fitting (NTfit). doi: 10.5281/zenodo.4104662
\bibitem{hapi} R.V. Kochanov, I.E. Gordon, L.S. Rothman, P. Wcislo, C. Hill, J.S. Wilzewski, HITRAN Application Programming Interface (HAPI): A comprehensive approach to working with spectroscopic data, J. Quant. Spectrosc. Radiat. Transfer 177, 15-30 (2016) DOI: 10.1016/j.jqsrt.2016.03.005
\end{thebibliography}
\end{document}